\documentclass[preprint,12pt]{elsarticle}

\usepackage[T1]{fontenc}
\usepackage{lmodern}

\usepackage{amsmath}
\usepackage{amssymb}

\usepackage{booktabs}
\usepackage{array}
\usepackage{longtable}
\usepackage{multirow}
\usepackage{float}

\graphicspath{{./}{./figures/}}

\usepackage{tikz}
\usepackage{pgfplots}
\pgfplotsset{compat=1.18}
\usetikzlibrary{arrows.meta,positioning,shapes.geometric,shapes.misc,fit,backgrounds,calc,patterns,patterns.meta}

\definecolor{snblue}{RGB}{38,38,38}     
\definecolor{snlblue}{RGB}{225,225,225} 
\definecolor{sngrey}{RGB}{106,106,106}  
\definecolor{sngreen}{RGB}{74,74,74}    
\definecolor{snorange}{RGB}{140,140,140}
\definecolor{snred}{RGB}{58,58,58}      

\usepackage{url}
\usepackage{xcolor}
\usepackage{textcomp}

\usepackage[protrusion=true,expansion=false]{microtype}

\usepackage{seqsplit}
\newcommand{\fname}[1]{\texttt{\seqsplit{#1}}}

\usepackage[colorlinks=false,allbordercolors={1 1 1}]{hyperref}
\usepackage{cleveref}

\journal{Journal of Systems and Software}

\begin{document}

\begin{frontmatter}

\title{Augment Engineering: A Methodology for Multi-Tool AI Orchestration Across Professional Domains}

\author[swift]{Elias Calboreanu\corref{cor1}}
\ead{ecalboreanu@theswiftgroup.com}
\ead[url]{https://orcid.org/0009-0008-9194-0589}

\cortext[cor1]{Corresponding author.}

\affiliation[swift]{organization={Swift North AI Lab, The Swift Group, LLC},
            state={Maryland},
            country={USA}}

\begin{abstract}
Organizations increasingly deploy separate purpose-built AI tools for code generation,
video production, presentation design, document processing, and project management,
often hiring domain specialists for each, recreating the staffing models AI was
expected to transform. Yet the meta-skills that make these tools effective, prompt
engineering (interaction-level optimization) and context engineering (structured input
pipeline design), are domain-portable: a practitioner who masters them can apply them
to any purpose-built AI tool in any domain.
This paper defines \emph{Augment Engineering} as the discipline of orchestrating
multiple purpose-built AI tools across distinct professional domains, applying prompt
and context engineering as portable competencies that transfer across tool boundaries.
We present a six-phase multi-tool orchestration methodology and four portability metrics.
A 5-month formative case study (November 2025--March 2026) documents a single practitioner
applying these skills across a ten-component orchestration stack (five purpose-built AI tools and five supporting infrastructure components) spanning seven
professional domains, producing work products that would traditionally involve separate
domain specialists. Two quantitative observations are consistent with the framework's
predictions: a Cochran-Armitage trend test ($n = 200$ interactions across two chat LLMs,
$p < 0.01$) shows first-pass acceptance rising with prompt-sophistication level, and a
Wright's Law fit ($n = 82$ non-excluded artifacts, $p < 0.01$) shows production
acceleration across the artifact portfolio. Because all observations come from a single practitioner, the inferential
statistics are exploratory: observations within one individual's workflow
are not fully independent, and the results should be interpreted as
hypothesis-generating rather than confirmatory. Neither test establishes portability across the
full 10-component / 7-domain portfolio; that claim awaits multi-practitioner replication.
Augment Engineering completes a three-discipline progression: Prompt Engineering
(one tool), Context Engineering (reproducible pipelines), Augment Engineering
(a portfolio of tools across domains).
\end{abstract}

\begin{highlights}
\item Augment Engineering orchestrates multiple AI tools across professional domains
\item Six-phase multi-tool orchestration methodology with defined phase criteria
\item Four portability metrics: transfer velocity, quality, overhead, coverage
\item Five-month single-practitioner case study: ten-component stack, seven domains
\item Prompt and context engineering transfer as portable cross-tool meta-skills
\end{highlights}


\end{frontmatter}

\section{Introduction}\label{sec:introduction}

Over a five-month period spanning November 2025 through March 2026, a single practitioner with no specialist training in video production, presentation design, curriculum design, academic publishing, or web deployment produced professional-grade deliverables across all five of these domains. During the same period, the practitioner also completed work in software engineering and contract proposal development, domains in which substantial expertise was held. The practitioner's toolkit comprised a ten-component orchestration stack distributed across seven professional domains, with no external team members serving as dedicated video editor, presentation designer, web developer, or curriculum specialist. The operational scope (one person, multiple tools, multiple domains) raises an immediate question: how was this possible? Is the underlying method reproducible, or does the outcome represent an artifact of idiosyncratic background, unusual talent, or fortunate circumstance?

The answer lies not in the raw power of the underlying AI systems, though that power is undeniably necessary. Rather, the competencies required to operate AI tools effectively, prompt engineering at the interaction level and context engineering~\cite{calboreanu2026context} at the pipeline level, are domain-portable: neither is tool-specific or domain-specific, and that portability is what makes the orchestration argument of this paper possible. A practitioner trained to structure inputs for Claude in order to draft a contract proposal can apply architecturally identical techniques to configure Gamma to produce a briefing deck, or HeyGen to generate training video content. The specific inputs differ markedly; the underlying methodology remains constant.

Two established disciplines currently address AI-integrated work at distinct levels of abstraction. Prompt engineering~\cite{white2023prompt, reynolds2021prompt} codifies techniques for optimizing discrete human-AI interactions, treating each query or task as an isolated optimization problem. Context engineering~\cite{calboreanu2026context} extends this to the pipeline level, prescribing structured input design that produces reproducible AI behavior across repeated invocations. Yet neither discipline directly addresses the cross-domain orchestration problem: how does a practitioner systematically coordinate multiple purpose-built AI tools across fundamentally different professional domains? How should outputs from one tool be structured as inputs to another? Where should quality gates be positioned at tool boundaries? By what metrics should coverage breadth be measured and optimized? These questions remain open. This paper addresses two research questions:

\begin{description}
\item[RQ1] Are prompt and context engineering skills portable across AI tool and professional domain boundaries, and if so, can the transfer be measured?
\item[RQ2] Can a systematic methodology for multi-tool orchestration be defined, codified, and illustrated through operational practice?
\end{description}

This work contributes in three ways. First, it establishes a formal definition of augment engineering as a distinct discipline, clearly delimited from prompt engineering, context engineering, human-AI teaming, multi-agent frameworks, and general AI adoption strategy, accompanied by a taxonomy of adjacent and overlapping concepts. Second, it presents a six-phase multi-tool orchestration methodology with explicitly defined inputs, outputs, and completion criteria for each phase, along with four quantifiable metrics for assessing cross-domain capability transfer and portfolio effectiveness. Third, it provides a formative case study grounded in a five-month operational study documenting a single practitioner's tool portfolio, cross-domain workflows, observed skill transfer patterns, and output quality metrics across seven professional domains using a ten-component orchestration stack (five purpose-built AI tools and five supporting infrastructure components), including domains in which the practitioner had received no prior specialist training. Statistical validation of the portability claim across practitioners and sites is explicitly future work (\Cref{sec:disc-future}).

\textit{Relationship to companion work.}
This paper is part of a research program that includes companion manuscripts on context engineering~\cite{calboreanu2026context}, governance architectures~\cite{calboreanu2026lattice, calboreanu2026trace}, autonomous task decomposition~\cite{calboreanu2026mandate}, and closed-loop software development~\cite{calboreanu2026closedloop}. The 200-interaction corpus analyzed here is the same corpus reported in the context engineering study, and the case study environment is the same Swift North AI Lab described in the companion works. What is \emph{new} in the present paper, and not reported elsewhere, is: (1)~the formal definition of augment engineering as a distinct discipline; (2)~the six-phase multi-tool orchestration methodology; (3)~the four portability metrics (transfer velocity, cross-domain output quality, orchestration overhead, and coverage breadth); and (4)~the cross-domain case study documenting skill transfer across seven professional domains and a ten-component orchestration stack. None of the companion works addresses cross-domain orchestration or proposes portability metrics.

The remainder of this paper proceeds as follows. \Cref{sec:related-work} surveys related work across AI-assisted development, multi-agent orchestration frameworks, human-AI collaboration models, prompt engineering, context engineering, workforce augmentation, and no-code/low-code development platforms, establishing the conceptual landscape and the positioning of augment engineering within it. \Cref{sec:framework} presents the augment engineering framework in detail, including the formal definition, the three-discipline progression, the six-phase orchestration methodology, the four portability metrics, and explicit scope boundaries. \Cref{sec:case-study} documents the Swift North AI Lab case study, presenting the tool portfolio inventory, domain-by-domain analysis with exemplar workflows, direct evidence of skill portability, and documented failure modes that illuminate the methodology's boundaries. \Cref{sec:discussion} examines limitations, generalizability constraints, the interaction between domain expertise and augment engineering skill, and directions for future work. \Cref{sec:conclusion} concludes.

\section{Related Work}\label{sec:related-work}

\noindent\textit{Note on reference maturity.} The multi-tool orchestration literature is in active preprint form; several scaffolding references below are arXiv preprints rather than peer-reviewed publications. Where possible, we cite published work; preprint status is noted in the reference list.

\subsection{AI-Assisted Software Development}\label{sec:rw-developer}

AI-assisted development has demonstrated substantial single-tool productivity gains: Copilot users completed coding tasks approximately 55\% faster~\cite{peng2023impact}, and autonomous agents such as SWE-Agent~\cite{yang2024sweagent} can resolve GitHub issues by iterating over repositories. Yet these evaluations measure single-tool performance on isolated coding tasks within a bounded domain. They do not address how practitioners orchestrate multiple tools across domain boundaries or whether competence with one AI tool generalizes to a different tool in a different professional domain.

\subsection{AI Tool Ecosystems and Multi-Agent Frameworks}\label{sec:rw-ecosystems}

Multi-agent frameworks such as LangChain~\cite{langchain2023}, AutoGPT~\cite{autogpt2023}, CrewAI~\cite{crewai2024}, and MetaGPT~\cite{hong2023metagpt} automate computational workflows by chaining LLM calls with memory and role differentiation. These frameworks operate within a single computational pipeline whose primitives are all LLM calls or LLM-derived function calls. Augment engineering addresses a different problem: the orchestration of heterogeneous, purpose-built tools (not all LLMs) by a practitioner who applies judgment at each tool boundary. The unit of orchestration in a multi-agent framework is the software pipeline; the unit in augment engineering is the individual practitioner.

\subsection{Human-AI Teaming}\label{sec:rw-teaming}

Human-AI collaboration research has established that complementary human-AI teams can exceed human-only performance~\cite{amershi2019guidelines, bansal2021does, lai2023towards}, with design guidelines for interaction, explanation, and graceful degradation. The conceptual unit in this literature, however, is a team of humans working with a single AI system. Augment engineering shifts both dimensions: the organizational unit is an individual practitioner, and the AI component is a portfolio of distinct tools. The concern is not decision accuracy within one system but the practitioner's ability to select, configure, and compose tools across domain boundaries.

\subsection{Prompt Engineering}\label{sec:rw-prompt}

Prompt engineering has matured into a learnable discipline with replicable techniques: pattern catalogs~\cite{white2023prompt}, prompt programming~\cite{reynolds2021prompt}, and chain-of-thought reasoning~\cite{wei2022chain} demonstrate that structured interaction design measurably improves model output. The underlying assumption, however, is that prompt optimization is tool-specific. Augment engineering claims a stronger proposition: that the mental models and practices that make a practitioner effective at structuring inputs for one AI tool transfer substantially to other tools, even those built on different architectures. This claim of cross-tool portability is not directly addressed in the prompt engineering literature surveyed here.

\subsection{Context Engineering}\label{sec:rw-context}

Context engineering~\cite{calboreanu2026context} defines the methodology for structuring AI input pipelines as five-role context packages, with 200 documented interactions showing that incomplete or poorly structured context drives 72\% of iteration cycles requiring human intervention. Augment engineering assumes that methodology as prerequisite: structured input design, role definition, and iterative refinement are the portable skills that carry across tools and domains. The question this paper takes up is what happens when those skills are applied across a tool portfolio rather than within a single tool. Context engineering is about the structure of inputs to any single tool; augment engineering is about selecting, adapting, and composing tools using context engineering as one core methodology.

Concurrent independent work by Vishnyakova~\cite{vishnyakova2026context} proposes a cumulative pyramid maturity model progressing from prompt engineering through context engineering to intent engineering and specification engineering. This convergence from separate research programs corroborates the discipline-progression concept: multiple groups have independently concluded that prompt engineering, context engineering, and higher-order orchestration skills form a cumulative hierarchy. The present framework differs in that it positions multi-tool orchestration (augment engineering) as the third level rather than intent or specification engineering, reflecting a practitioner-oriented rather than enterprise-architecture perspective.

\subsection{Workforce Augmentation and Automation Economics}\label{sec:rw-workforce}

Automation economics operates at the macro level, modeling how labor markets adjust to technological change~\cite{autor2015there, acemoglu2019automation} and measuring aggregate productivity shifts from generative AI~\cite{brynjolfsson2023generative}. Augment engineering contributes at the micro level: it describes the mechanisms by which an individual practitioner orchestrates multiple tools. The two perspectives are complementary but distinct.

\subsection{No-Code and Low-Code Platforms}\label{sec:rw-nocode}

No-code/low-code platforms~\cite{sahay2020supporting} automate data flows between tools but do not address the practitioner's cognitive work at tool boundaries: validating outputs, applying domain judgment, and deciding whether a tool's output is suitable for downstream consumption. Augment engineering sits above workflow automation, concerned with how practitioners think about and compose tools, not how those tools are computationally integrated.

\begin{table*}[!htbp]
\caption{Comparison of augment engineering with related approaches. Each row identifies
the field, its primary focus, what it measures, and the gap that augment engineering
addresses. The central gap: few works directly address the practitioner-method question of how a single
practitioner systematically orchestrates multiple purpose-built AI tools across professional domains
using portable prompt and context engineering skills.}
\label{tab:related-approaches}
\centering
\small
\resizebox{\textwidth}{!}{%
\begin{tabular}{@{}p{3.0cm}p{3.2cm}p{3.6cm}p{5.0cm}@{}}
\toprule
\textbf{Field} & \textbf{Primary Focus} & \textbf{What It Measures} & \textbf{Gap Augment Engineering Fills} \\
\midrule
AI-Assisted Development
  & Single-tool coding productivity
  & Task completion time, code quality per tool
  & Single tool, single domain; does not address multi-tool orchestration or skill portability \\
\addlinespace
Multi-Agent Frameworks
  & LLM call chaining in software pipelines
  & Pipeline throughput, task success rate
  & Automates LLM pipelines; does not orchestrate heterogeneous tools across professional domains \\
\addlinespace
Human-AI Teaming
  & Team-AI collaboration dynamics
  & Decision accuracy, trust calibration
  & Studies teams with one AI system; does not address one practitioner with many AI tools \\
\addlinespace
Prompt Engineering
  & Individual interaction optimization
  & Output quality, accuracy, format compliance
  & Optimizes within one tool; does not address skill portability across tool boundaries \\
\addlinespace
Context Engineering
  & Structured input pipeline design
  & Reproducibility, first-pass acceptance
  & Structures inputs for one tool; does not address cross-domain orchestration \\
\addlinespace
Automation Economics
  & Macro labor market effects
  & Employment, wages, GDP impact
  & Analyzes markets; does not prescribe practitioner-level orchestration methodology \\
\addlinespace
No-Code/Low-Code
  & Workflow automation
  & Data flow integration, process automation
  & Automates data flows; does not address practitioner skill transfer or domain judgment \\
\bottomrule
\end{tabular}%
}
\end{table*}

\subsection{Summary of Gaps}\label{sec:rw-summary}

\Cref{tab:related-approaches} synthesizes the positioning. The central gap: existing work does not directly address the practitioner-method question: how a single practitioner systematically orchestrates multiple purpose-built AI tools across professional domains using portable prompt and context engineering skills. Augment engineering addresses this gap by elevating tool orchestration from ad-hoc practice to methodological discipline.

\section{The Augment Engineering Framework}\label{sec:framework}

This section defines augment engineering as a discipline, establishes its scope boundaries
relative to adjacent fields, presents the multi-tool orchestration methodology, and
formalizes the portability metrics by which cross-domain capability is assessed.

\subsection{Definition and Scope}\label{sec:definition}

\begin{quote}
\textbf{Augment Engineering} is the discipline of orchestrating multiple purpose-built
AI tools across distinct professional domains, applying prompt engineering and context
engineering skills as portable competencies that transfer across tool boundaries, to
enable a single practitioner to produce output spanning roles that traditionally require
separate domain specialists.
\end{quote}

The term ``engineering'' is deliberate: augment engineering prescribes repeatable processes
with measurable outcomes, not strategic guidance or organizational readiness assessments.
An augment engineer designs cross-domain workflows the same way a systems engineer designs
integration architectures: with defined interfaces, quality gates at boundaries, structured
handoff patterns, and iterative optimization based on observed results.

The central claim is not that AI tools are powerful; that is necessary but not sufficient.
The claim is that prompt engineering and context engineering are \emph{transferable
meta-skills}. A practitioner who understands how to structure interactions with AI systems
(prompt engineering) and how to build reproducible input pipelines (context engineering)
can apply those skills to any purpose-built AI tool in any professional domain. The
augment engineer does not need to be a video production specialist to produce professional
video content. They do not need to be a graphic designer to produce professional
presentations. They need to understand how to structure inputs, validate outputs, and
integrate tool outputs into a coherent cross-domain workflow.

The scope of augment engineering is bounded by three exclusions.
First, it does not prescribe how to write effective prompts; that is the domain of
prompt engineering.
Second, it does not prescribe how to structure input data for reproducible model behavior;
that is the domain of context engineering~\cite{calboreanu2026context}.
Third, it does not prescribe enterprise AI strategy, vendor selection criteria, or
organizational change management; those are the domains of AI adoption and digital
transformation frameworks.
Augment engineering assumes that prompt and context engineering competencies exist within
the practitioner, that multiple purpose-built AI tools are available, and that the
objective is to orchestrate those tools across professional domains to produce
cross-domain operational capability.

Three terms recur throughout this paper and are used consistently. \emph{Augment
engineering} is the discipline defined above. The \emph{augment engineering methodology}
is the six-phase orchestration process (\Cref{sec:methodology}) that operationalizes that
discipline. The \emph{augment engineering framework} denotes the methodology together
with the portability metrics (\Cref{sec:metrics}) and the governance-checkpoint principle
(\Cref{sec:governance}) developed in this section. In short, the discipline names the
field; the methodology and the framework are this paper's contributions to it.

\subsection{The Three-Discipline Progression}\label{sec:hierarchy}

\begin{figure}[t]
  \centering
  \begin{tikzpicture}[
    font=\sffamily,
    tier/.style={
      rectangle, rounded corners=3pt, draw=black!70, line width=0.6pt,
      minimum width=7.2cm, minimum height=1.55cm,
      align=center, inner sep=5pt
    },
    tier1/.style={tier, fill=black!6},
    tier2/.style={tier, fill=black!14},
    tier3/.style={tier, fill=black!26, line width=0.9pt},
    scopecap/.style={font=\footnotesize\sffamily, align=left, text width=3.4cm},
    up/.style={-{Stealth[length=2.3mm]}, line width=0.8pt, black!75},
  ]
    \node[tier1] (t1) at (0, 0) {%
      \textbf{Prompt Engineering}\\[1pt]
      \footnotesize Interact with \emph{one} tool for one task};
    \node[tier2] (t2) at (0, 1.95) {%
      \textbf{Context Engineering}\\[1pt]
      \footnotesize Structure inputs for \emph{reproducible} behavior};
    \node[tier3] (t3) at (0, 3.90) {%
      \textbf{Augment Engineering}\\[1pt]
      \footnotesize Orchestrate a \emph{portfolio} of tools across domains};

    \draw[up] (t1.north) -- node[right=1pt, font=\scriptsize\itshape]{enables} (t2.south);
    \draw[up] (t2.north) -- node[right=1pt, font=\scriptsize\itshape]{enables} (t3.south);

    \node[scopecap, anchor=west] at (4.0, 0)    {\textit{Scope:} one tool\\ \textit{Unit:} prompt};
    \node[scopecap, anchor=west] at (4.0, 1.95) {\textit{Scope:} one tool, many runs\\ \textit{Unit:} context pipeline};
    \node[scopecap, anchor=west] at (4.0, 3.90) {\textit{Scope:} many tools, many domains\\ \textit{Unit:} orchestrated workflow};

    \node[anchor=north, font=\scriptsize\itshape, text width=11cm, align=center]
      at (1.7, -1.15)
      {Each level is cumulative: augment engineering requires competence at both lower levels.};
  \end{tikzpicture}
  \caption{The three-discipline progression. Each level builds on the
  portable skills developed at the level below. Prompt engineering skills
  enable context engineering; context engineering skills enable augment
  engineering. The progression is cumulative: augment engineering requires
  competence at both lower levels. Ordering matters for workforce
  development sequencing.}
  \label{fig:discipline-hierarchy}
\end{figure}
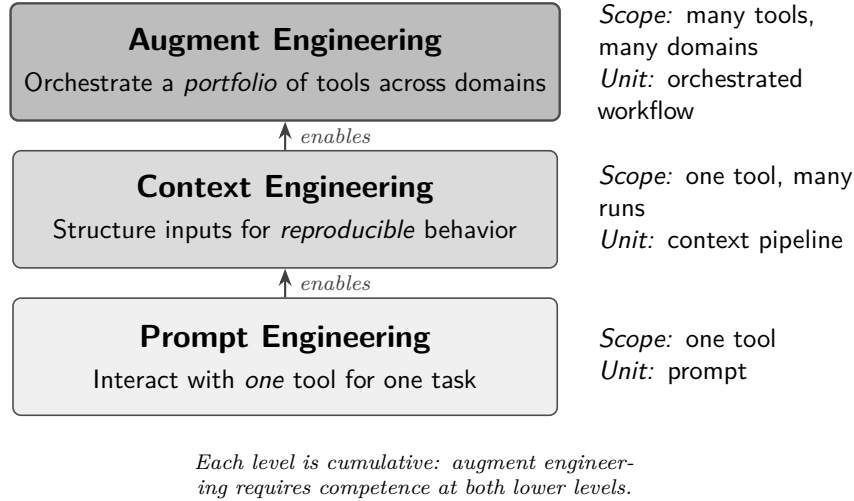

\begin{table}[t]
\caption{Comparison of augment engineering with adjacent concepts. Each concept operates
at a distinct level and addresses a distinct question. Augment engineering is the only
concept that addresses multi-tool, multi-domain orchestration by a single practitioner.}
\label{tab:concept-comparison}
\centering
\small
\begin{tabular}{@{}lp{2.2cm}p{3cm}p{3.5cm}@{}}
\toprule
\textbf{Concept} & \textbf{Level of Operation} & \textbf{Core Question} & \textbf{Output} \\
\midrule
Prompt Engineering & Single tool, single task & How do I get the best output from this interaction? & Optimized individual output \\
\addlinespace
Context Engineering & Single tool, structured input & How do I make this tool's behavior reproducible? & Reproducible model behavior \\
\addlinespace
Augment Engineering & Multiple tools, multiple domains & How do I orchestrate a portfolio of AI tools across domains? & Cross-domain operational capability \\
\addlinespace
AI Adoption & Enterprise strategy & Should we adopt AI? & Adoption roadmaps \\
\addlinespace
Human-AI Teaming & Team dynamics & How do humans and AI collaborate? & Collaboration frameworks \\
\bottomrule
\end{tabular}
\end{table}

Augment engineering completes a three-level discipline progression for AI-integrated work.
Each level operates at a distinct scope, addresses a distinct question, and produces
a distinct output. \Cref{tab:concept-comparison} summarizes the distinctions;
\Cref{fig:discipline-hierarchy} illustrates the layered relationship.

\textbf{Level 1: Prompt Engineering} optimizes single interactions within a single tool. \textbf{Level 2: Context Engineering}~\cite{calboreanu2026context} structures input pipelines (authority documents, reference materials, templates, conductor prompts) so that a tool's behavior becomes reproducible across sessions and operators. Both levels are well-established; their definitions are summarized in \Cref{tab:concept-comparison}.

\textbf{Level 3: Augment Engineering} is the novel contribution. The practitioner's objective shifts from operating one tool well to orchestrating a portfolio of tools across professional domains. This requires treating prompt and context engineering skills as portable competencies applied at each tool boundary: the same structured-input discipline that produces reproducible Claude behavior for a contract proposal also structures Gamma inputs for a briefing deck and HeyGen inputs for training video. The orchestration challenge (quality gates, format translation, governance checkpoints, cross-tool iteration) is what distinguishes augment engineering from applying context engineering to multiple tools independently.

The progression is cumulative: a practitioner who cannot write effective prompts will not produce reliable context packages, and a practitioner who cannot produce reliable context packages will not sustain effective multi-tool orchestration. This dependency has direct implications for workforce development sequencing.

\subsection{The Multi-Tool Orchestration Methodology}\label{sec:methodology}

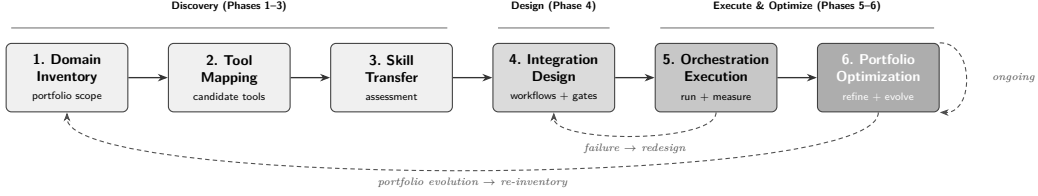
\begin{figure*}[t]
  \centering
  \resizebox{\textwidth}{!}{%
  \begin{tikzpicture}[
    font=\sffamily,
    phase/.style={
      rectangle, rounded corners=3pt, draw=black!75, line width=0.6pt,
      minimum width=2.4cm, minimum height=1.35cm,
      align=center, inner sep=3pt, font=\scriptsize\sffamily
    },
    discovery/.style={phase, fill=black!6},
    design/.style={phase, fill=black!14},
    execute/.style={phase, fill=black!22},
    optimize/.style={phase, fill=black!32, text=white},
    arr/.style={-{Stealth[length=2mm]}, line width=0.75pt, black!80},
    fb/.style={-{Stealth[length=1.8mm]}, line width=0.6pt, black!70,
               dash pattern=on 2.5pt off 1.8pt},
    band/.style={font=\scriptsize\sffamily\bfseries, align=center},
  ]
    \def\xs{3.2}
    \node[discovery] (p1) at (0*\xs, 0) {\textbf{1.~Domain}\\\textbf{Inventory}\\[1pt]\tiny portfolio scope};
    \node[discovery] (p2) at (1*\xs, 0) {\textbf{2.~Tool}\\\textbf{Mapping}\\[1pt]\tiny candidate tools};
    \node[discovery] (p3) at (2*\xs, 0) {\textbf{3.~Skill}\\\textbf{Transfer}\\[1pt]\tiny assessment};
    \node[design]    (p4) at (3*\xs, 0) {\textbf{4.~Integration}\\\textbf{Design}\\[1pt]\tiny workflows + gates};
    \node[execute]   (p5) at (4*\xs, 0) {\textbf{5.~Orchestration}\\\textbf{Execution}\\[1pt]\tiny run + measure};
    \node[optimize]  (p6) at (5*\xs, 0) {\textbf{6.~Portfolio}\\\textbf{Optimization}\\[1pt]\tiny refine + evolve};

    \foreach \a/\b in {p1/p2, p2/p3, p3/p4, p4/p5, p5/p6} {
      \draw[arr] (\a.east) -- (\b.west);
    }

    \node[band, anchor=south, font=\tiny\sffamily\bfseries] at (1*\xs, 1.15)   {\textsc{Discovery} (Phases 1--3)};
    \node[band, anchor=south, font=\tiny\sffamily\bfseries] at (3*\xs, 1.15)   {\textsc{Design} (Phase 4)};
    \node[band, anchor=south, font=\tiny\sffamily\bfseries] at (4.5*\xs, 1.15) {\textsc{Execute \& Optimize} (Phases 5--6)};

    \draw[black!55, line width=0.6pt] (-1.1, 0.98) -- ({2*\xs+1.1}, 0.98);
    \draw[black!55, line width=0.6pt] ({3*\xs-1.1}, 0.98) -- ({3*\xs+1.1}, 0.98);
    \draw[black!55, line width=0.6pt] ({4*\xs-1.1}, 0.98) -- ({5*\xs+1.1}, 0.98);

    \draw[fb] (p5.south) .. controls ({4*\xs}, -1.3) and ({3*\xs}, -1.3) .. (p4.south)
      node[midway, below, font=\tiny\itshape, text=black!70]{failure $\rightarrow$ redesign};
    \draw[fb] (p6.south) .. controls ({5*\xs}, -2.2) and ({0*\xs}, -2.2) .. (p1.south)
      node[pos=0.5, below=0pt, font=\tiny\itshape, text=black!70]{portfolio evolution $\rightarrow$ re-inventory};

    \draw[fb] (p6.north east) .. controls ({5*\xs+1.9}, 0.7) and ({5*\xs+1.9}, -0.7) .. (p6.south east);
    \node[font=\tiny\itshape, anchor=west, text=black!70] at ({5*\xs+2.1}, 0) {ongoing};
  \end{tikzpicture}%
  }
  \caption{The six-phase multi-tool orchestration methodology. Phases 1--3
  (Discovery) identify the practitioner's domain landscape and assess skill
  transferability. Phase~4 (Design) specifies cross-tool workflows with
  quality gates at tool boundaries. Phase~5 (Execution) operates the
  portfolio and collects portability metrics. Phase~6 optimizes the
  portfolio based on execution data. Dashed arrows indicate feedback
  loops: failure cases refine integration designs, and portfolio evolution
  triggers re-assessment. Phase~6 recurs periodically; it does not terminate.}
  \label{fig:orchestration-methodology}
\end{figure*}

The three-discipline progression establishes \emph{what} augment engineering is;
the methodology establishes \emph{how} it is practiced.
The core engineering contribution is a six-phase orchestration methodology.
Each phase produces defined artifacts, has measurable completion criteria,
and feeds into the subsequent phase. \Cref{fig:orchestration-methodology} illustrates
the methodology structure.

\subsubsection{Phase 1: Domain Inventory}\label{sec:phase1}

The first phase identifies the professional domains required by the practitioner's
role or the organization's mission. The practitioner catalogs every professional
domain in which they must produce work products, treating a ``domain'' as a field that
traditionally requires specialist training: software development, technical writing,
video production, presentation design, data analysis, curriculum design, project
management, and so on. The inventory is exhaustive; the purpose is to identify the
full breadth of domains across which the practitioner operates or could operate. The
phase produces a domain inventory listing every domain, the work products required in
each, and the current method of production (specialist hire, outsource, in-house
non-specialist, or unaddressed), and completes when every deliverable type in the
practitioner's portfolio has been mapped to a domain.

\subsubsection{Phase 2: Tool Mapping}\label{sec:phase2}

The second phase takes the domain inventory from Phase~1 and identifies the best
available purpose-built AI tool for each domain. For each domain, the practitioner
evaluates candidate tools against three criteria: whether the tool accepts structured
inputs that prompt and context engineering skills can optimize, whether it produces
outputs of sufficient quality to meet professional standards in the target domain, and
whether its output format is compatible with the practitioner's cross-domain workflow
(i.e., whether outputs can serve as inputs to tools in adjacent domains).

Not every domain will have a suitable AI tool. The tool mapping phase explicitly
identifies domains where no adequate tool exists, where the available tools do not
accept structured inputs amenable to prompt and context engineering, or where the
quality threshold cannot be met. These domains remain outside the augmentation
portfolio until the tool landscape evolves. The phase produces a tool-domain mapping
table listing each domain, the selected AI tool (or ``no suitable tool''), and the
rationale for each selection; it completes when every domain in the inventory has been
either mapped to a tool or documented as unmappable with stated reasons.

\subsubsection{Phase 3: Skill Transfer Assessment}\label{sec:phase3}

The third phase assesses how prompt and context engineering skills apply to each
tool in the portfolio. For each tool-domain pairing, the practitioner determines which
techniques transfer directly, which require adaptation, and which do not apply.
Transfer assessment addresses four dimensions:

\begin{enumerate}
  \item \textbf{Input structuring}: Can the tool accept structured inputs (authority
        documents, templates, constraints, rubrics) analogous to context packages
        built for other tools?
  \item \textbf{Output validation}: Can the practitioner apply the same quality
        validation methodology (rubric-based assessment, iterative refinement)
        used with other tools?
  \item \textbf{Iteration patterns}: Does the tool support the same iterative
        refinement workflow (generate, evaluate, revise) that prompt engineering
        skills enable?
  \item \textbf{Integration compatibility}: Can the tool's outputs serve as
        structured inputs to other tools in the portfolio without manual
        reformatting?
\end{enumerate}

The phase produces a skill transfer matrix documenting, for each tool-domain pairing,
which skills transfer directly, which require adaptation, and what domain-specific
knowledge (if any) the practitioner must acquire. It completes when every tool-domain
pairing has a documented transfer assessment across all four dimensions.

\subsubsection{Phase 4: Integration Design}\label{sec:phase4}

The fourth phase designs the cross-tool workflow: how outputs from one tool become
inputs to another, and where quality gates are placed at tool boundaries. Drawing on
the skill transfer matrix from Phase~3 and the tool-domain mapping from Phase~2, the
practitioner designs end-to-end workflows that span multiple tools and domains. For
each workflow, the integration design specifies which tools contribute at each stage,
the handoff format between tools, where quality gates sit at tool boundaries, and what
fallback procedures apply when a tool's output fails the quality gate.

Integration design is where context engineering skills become essential at the
orchestration level. Each tool boundary is effectively a context engineering problem:
the output of Tool~A must be structured as a valid input for Tool~B, with the same
attention to authority, reference material, and constraints that govern any context
package~\cite{calboreanu2026context}.

The integration design also specifies \emph{governance checkpoints}: points in the
cross-tool workflow where human review is required before outputs propagate to the
next tool. Each tool boundary is a natural governance checkpoint where output quality
must be validated before becoming input to the next stage. This connects augment
engineering to the broader question of architectural trust: confidence in the
multi-tool pipeline derives from the integration design and governance checkpoints,
not from trust in any individual AI tool, a principle developed in this author's work
on governance-first architectures~\cite{calboreanu2026lattice}. The phase produces
cross-domain workflow specifications for each multi-tool pipeline (tool sequence,
handoff formats, quality gates, governance checkpoints, and fallback procedures) and
completes when each multi-tool workflow has a documented integration design that has
been reviewed and approved for execution.

\subsubsection{Phase 5: Orchestration Execution}\label{sec:phase5}

The fifth phase operates the portfolio, producing cross-domain deliverables using
the integration designs from Phase~4. The practitioner executes the designed
workflows and, during execution, collects data on the four portability metrics defined
in \Cref{sec:metrics}: transfer velocity, cross-domain output quality, orchestration
overhead, and coverage breadth.

Orchestration execution also surfaces failure cases: domains where the skill transfer
assessment was overly optimistic, tools where output quality does not meet
professional standards despite structured inputs, and workflow boundaries where
handoff friction is higher than anticipated. These failure cases are informative;
they refine the portfolio and the integration designs. The phase produces
cross-domain work products, portability metric data, and a logged set of failure
cases with documented causes, and completes when the practitioner has executed at
least one full cycle of each designed multi-tool workflow and collected portability
metric data.

\subsubsection{Phase 6: Portfolio Optimization}\label{sec:phase6}

The sixth phase iterates on the tool portfolio and integration designs based on
the metrics and failure case logs from Phase~5. The practitioner analyzes execution
results to identify optimization opportunities across three dimensions: tool
substitution (replacing underperforming tools with alternatives that better accept
structured inputs or produce higher-quality outputs), workflow redesign (restructuring
cross-tool pipelines to reduce handoff friction or improve quality gate
effectiveness), and portfolio expansion (adding new tool-domain pairings as the
practitioner's skill transfer competence grows or as the tool landscape evolves).

Portfolio optimization is continuous. As new AI tools emerge, as existing tools
gain capabilities, and as the practitioner accumulates cross-domain experience,
the portfolio evolves. The augment engineer maintains the domain inventory, tool
mapping, and integration designs as living documents, not static artifacts. The phase
produces revised tool-domain mappings, updated integration designs, and updated
portability metrics reflecting optimization changes; a cycle completes when the
practitioner has reviewed all failure cases from Phase~5 and documented optimization
decisions with rationale. This phase recurs periodically; it does not terminate.

\subsection{Portability Metrics}\label{sec:metrics}

To assess whether portfolio optimization is effective and whether the
methodology produces measurable cross-domain capability, augment engineering
defines a set of \emph{portability metrics}.
These metrics measure the effectiveness of skill transfer across tool
boundaries and domains. They are distinct from productivity metrics that
compare output volume against staffing baselines; portability metrics instead
measure the mechanism that enables cross-domain operation: the transferability
of prompt and context engineering skills.

We define four portability metrics, grounding each in established mathematical frameworks from learning curve theory~\cite{hutter2021learning,viering2021shape}, coordination cost modeling~\cite{kim2025scaling}, and human-AI collaboration evaluation~\cite{fragiadakis2024evaluating}. Throughout this section, the following notation is used: $n$ denotes the cumulative count of AI tools adopted by a practitioner; $d$ indexes a professional domain from the practitioner's coverage set $D$; $T_0$ is the baseline time-to-proficiency measured on the first tool; $T_v(n)$ is the expected time-to-proficiency on the $n$-th tool; $\beta > 0$ is the dimensionless skill-transfer acceleration exponent (analogous to the Wright's Law learning-rate exponent); $Q_d \in \{0,1\}$ is a binary quality indicator for domain $d$; $O_h \in [0,1]$ is the orchestration-overhead ratio; and $C_b$ is the coverage-breadth count. All time-valued quantities are expressed in hours unless otherwise noted.

\begin{enumerate}
  \item \textbf{Transfer velocity} ($T_v$): The time from first use of a new
        AI tool to productive output that meets professional standards in the
        target domain. The claim is that practitioners with strong prompt and context
        engineering foundations achieve productive use of new tools within hours
        or days rather than weeks, because the interaction patterns, input
        structuring methods, and output validation techniques transfer even
        when the tool interface is unfamiliar.
        Formally, transfer velocity follows a power-law learning curve~\cite{viering2021shape}:
        \begin{equation}\label{eq:tv}
          T_v(n) = T_0 \cdot n^{-\beta}, \quad \beta > 0
        \end{equation}
        where $T_0$ is the baseline time-to-proficiency with the first tool, $n$ is the
        cumulative number of tools adopted, and $\beta$ is the skill transfer
        acceleration exponent. This formulation is the practitioner-level analog
        of Wright's law~\cite{wright1936factors, narayanan2025wrights}, which predicts that unit cost
        decreases as a power function of cumulative production experience.
        Millinghoffer et al.~\cite{millinghoffer2025transfer} independently demonstrate
        that transfer effects across related tasks follow analogous power-law learning
        curves in multi-task settings, providing theoretical support for this functional
        form at the task level.
        A positive $\beta$ implies each successive tool requires less ramp-up time;
        $\frac{dT_v}{dn} < 0$ indicates accelerating transfer. The empirical
        six-adoption fit of this equation ($R^2 = 0.29$, $p > 0.10$, $n = 6$) is directionally consistent with the Wright's Law curve in \Cref{fig:wrights-law} but not statistically powered. These six observations are per-domain-entry events, not independent per-tool observations, so this result does not test the per-tool power law; it is reported as illustrative only.

  \item \textbf{Cross-domain output quality} ($Q_d$): An assessment
        of whether the practitioner's output in a domain where they have no specialist
        training meets the professional standards of that domain.
        Formally:
        \begin{equation}\label{eq:qd}
          Q_d(d) \in \{0, 1\}
        \end{equation}
        where $Q_d = 1$ indicates validated acceptance through domain-appropriate
        external review: advancement past desk-review to substantive peer review for academic publications,
        stakeholder acceptance for contract deliverables, production deployment for
        software, and client acceptance for presentations and training materials.
        This binary formulation parallels transferability metrics such as the
        F-OTCE score~\cite{tan2022transferability}, which quantifies how much a source
        model benefits target task learning. Celis et al.~\cite{celis2025mathematical}
        provide independent theoretical support for threshold-based quality metrics:
        their mathematical framework for AI-human integration demonstrates that job
        success probability exhibits sharp phase transitions as subskill abilities
        cross critical thresholds, justifying a binary rather than continuous quality
        measure at the domain level.
        Ordinal measurement (e.g., revision round counts, reviewer scores, number of
        audit passes before acceptance) would add granularity and is proposed as future work.

  \item \textbf{Orchestration overhead} ($O_h$): The proportion of total work time
        spent coordinating between tools (formatting handoffs, managing quality gates,
        resolving integration failures) versus producing within tools.
        Formally:
        \begin{equation}\label{eq:oh}
          O_h = \frac{t_{\text{coord}}}{t_{\text{coord}} + t_{\text{prod}}}
        \end{equation}
        where $t_{\text{coord}}$ is coordination time (format conversion, quality gate
        review, integration debugging) and $t_{\text{prod}}$ is in-tool production time,
        both estimated from timestamped workflow records where available (Git commits, Jira lane
        transitions, deployment logs) or reconstructed post-hoc from event logs where continuous sub-task time tracking was not instrumented. Precision depends on data-collection granularity.
        Lower overhead indicates more effective integration design.
        Research on multi-agent coordination confirms that overhead scales with
        interaction depth and can be reduced through centralized
        orchestration~\cite{kim2025scaling,dang2025evolving}; specifically, uniform
        multi-tool pipelines over-process simple tasks, and difficulty-aware
        orchestration reduces unnecessary overhead~\cite{su2025difficulty}.
        In the present study, we observe $\frac{dO_h}{dA} < 0$: overhead decreases
        as the degree of automation $A$ at tool boundaries increases (e.g., pipelines with Jira-as-Code automation (product development, $O_h \approx 8\%$) operate below pipelines without automated authority transfer (training content, $O_h \approx 18\%$)).

  \item \textbf{Coverage breadth} ($C_b$): The number of distinct professional domains
        in which a single practitioner produces work products that meet professional
        standards.
        Formally:
        \begin{equation}\label{eq:cb}
          C_b(t) = |\{d : Q_d(d) = 1\}|
        \end{equation}
        Coverage breadth is the most direct measure of augment engineering
        effectiveness: a practitioner who covers 7 professional domains with validated
        output quality demonstrates broader augmentation than one who covers 3.
        Workforce usage data indicate that LLMs are in active use for at least 25\% of tasks across approximately
        36\% of occupations~\cite{shao2025future}; $C_b$ operationalizes this
        breadth at the individual practitioner level.
\end{enumerate}

These metrics form a coupled system. The optimization objective of augment engineering
can be stated as:
\begin{equation}\label{eq:optimization}
  \max_{D_{\text{op}} \subseteq D_{\text{candidate}}} \; |D_{\text{op}}| \quad \text{subject to} \quad Q_d(d) = 1 \;\; \forall \, d \in D_{\text{op}}, \quad O_h \leq \tau
\end{equation}
where $D_{\text{candidate}}$ is the set of professional domains the practitioner could
potentially enter, $D_{\text{op}}$ is the operational subset the practitioner actually
serves, and $\tau$ is the maximum tolerable overhead ratio. The objective expresses
the augment-engineering goal: enter as many domains as possible while sustaining
the quality threshold on each and keeping coordination cost below the practitioner's
tolerance. A practitioner with high transfer velocity
but low cross-domain output quality is learning tools quickly but not producing usable
results. A practitioner with high coverage breadth but high orchestration overhead
($O_h > \tau$) is covering many domains but spending excessive time on coordination
rather than production. Effective augment engineering maximizes coverage breadth
while maintaining quality and keeping overhead below the practitioner's tolerance.

\subsection{Orchestration Patterns}\label{sec:patterns}

Multi-tool orchestration generates recurring structural patterns at tool boundaries. The six patterns below form an inductive taxonomy abstracted from the four case-study pipelines (\Cref{sec:cs-workflows}); they are descriptive categories derived from observation, not independently validated constructs:
(1)~\textit{authority-document chain}: an authority document from one tool becomes the governing constraint for the next;
(2)~\textit{format translation}: output from one tool is restructured to match the input schema of another (e.g., Claude markdown to LaTeX);
(3)~\textit{quality-gate handoff}: human validation at a tool boundary before output propagates downstream;
(4)~\textit{governance checkpoint}: organizational review embedded in the workflow (audit pass, compliance check);
(5)~\textit{terminal-node placement}: tools whose outputs cannot serve as structured inputs (e.g., HeyGen video) are positioned at workflow endpoints;
(6)~\textit{automated authority transfer}: Jira-as-Code or CI/CD pipelines that propagate context across tool boundaries without manual reformatting.
Integration design (\Cref{sec:phase4}) selects cross-tool pairings where handoff formats are compatible, governance checkpoints are feasible, and domain requirements justify the coordination cost. As Xu et al.~\cite{xu2026evolution} observe, multi-tool orchestration shifts the decision space from binary tool selection to coupled decisions involving dependency modeling and scheduling under constraints.

\subsection{Governance Integration}\label{sec:governance}

Augment engineering does not operate in isolation from organizational governance.
The integration designs produced in Phase~4 encode governance decisions: where human
review is required, what quality thresholds trigger rejection at tool boundaries,
and what fallback procedures exist when a tool's output is rejected.

This governance dimension connects augment engineering to the broader architectural
trust thesis articulated in this author's prior work on governance-first
architectures~\cite{calboreanu2026lattice} and machine-readable specification
frameworks~\cite{calboreanu2026mandate}.
At the orchestration level, the trust question becomes: ``Do we trust this multi-tool
pipeline?'' rather than ``Do we trust this individual AI tool?''
The former is answerable through engineering validation (integration design review,
quality gate effectiveness, cross-domain output quality assessment); the latter is not.

Each tool boundary in a multi-tool workflow is a natural governance checkpoint.
When the output of a code generation tool becomes the input to a deployment pipeline,
when the output of a writing tool becomes the input to a presentation tool, when the
output of a curriculum design tool becomes the input to a video production tool,
each boundary is a point where quality must be validated and governance must be applied.
Organizations that orchestrate multiple AI tools without designing governance into
the tool boundaries risk compounding errors across the pipeline: a quality failure in
an upstream tool propagates through downstream tools, potentially amplifying the
original error at each boundary.

Effective augment engineering therefore requires that governance checkpoints be designed
as first-class components of the integration, not afterthoughts applied to constrain
the workflow after it is already operational.

\subsection{Scope Boundaries}\label{sec:scope-boundaries}

To prevent misapplication and scope creep, it is important to state what augment
engineering is \emph{not}.

\textbf{Augment engineering is not a tools review.} The paper does not evaluate which
AI tools are ``best'' for any domain. The tool portfolio presented in the case study
(\Cref{sec:case-study}) is one instantiation of the methodology; it is not the
contribution. Different practitioners in different organizations will assemble
different portfolios. The contribution is the orchestration methodology, not the
specific tools.

\textbf{Augment engineering is not a productivity study.} The paper does not measure
``hours saved'' or ``cost reduction.'' Portability metrics (\Cref{sec:metrics}) measure
the mechanism of skill transfer across tool boundaries, not staffing efficiency.
The question is ``can one practitioner cover multiple professional domains?'' not
``is one practitioner cheaper than a team?''

\textbf{Augment engineering is not a staffing analysis.} The paper does not compare
augmented headcount against conventional headcount using labor category baselines.
That framing reduces the concept to staffing optimization, which mischaracterizes
the contribution. The contribution is a practitioner-level engineering methodology
for cross-domain orchestration.

\textbf{Augment engineering is not a curriculum evaluation.} The methodology has been codified into a structured instructional format, but this paper does not evaluate learner outcomes. Curriculum effectiveness evaluation is a separate study with distinct research questions and data requirements.

\section{Case Study: Multi-Tool Orchestration at Swift North AI Lab}\label{sec:case-study}

\subsection{Lab Environment and Study Design}\label{sec:cs-environment}

The study was conducted at Swift North AI Lab, a division of The Swift Group, LLC, operating on a purpose-built AI research cluster of four Mac mini M4 Pro workstations. The study period spans five months, from November 2025 through March 2026, during which the lab maintained continuous operational records under a formal Zero Trust governance framework with zone-based containment, kill switch mechanisms, and mandatory human oversight gates. All work was tracked through a Jira-based project management system configured with a seven-lane automated development pipeline. These lab conditions (purpose-built infrastructure, governance-first operating discipline, instrumented workflows, and five months of continuous record-keeping) are the enabling success factors for the methodology under study, and the case study is structured so that any practitioner operating under comparable conditions could in principle reproduce the orchestration patterns documented here.

This is a longitudinal study of one individual operating across multiple professional domains using a portfolio of AI tools, not a team productivity study. The case-study practitioner operated from a software engineering and cybersecurity baseline, where the prompt and context engineering competencies applied throughout this study were originally developed, and had no prior professional experience in video production, presentation design, web deployment, curriculum design, or academic publishing, the domains that figure prominently in the study's outcome portfolio. All cross-domain work presented here was conducted by applying portable prompt and context engineering skills to purpose-built AI tools, without prerequisite domain expertise.

\subsection{Output Quantification Methodology}\label{sec:cs-counting}

Quantitative claims in this section draw on three classes of evidence with differing precision. \textbf{Instrumented counts} (artifact totals, test pass rates, deployment counts, Jira lane transitions, Git commit history) are extracted from system records and treated as precise. \textbf{Reconstructed estimates} (orchestration overhead $O_h$, transfer velocity $T_v$ outside the Jira-tracked software pipeline) are derived post hoc from event-log timestamps and practitioner time logs and are reported as approximate ratios or order-of-magnitude figures. \textbf{Coded interaction data} (prompt-sophistication levels, first-pass acceptance) come from a single practitioner-coder applying the rubric in \ref{app:rubric}; inter-rater reliability was not assessed. Counting rules and data sources for each class are documented below to support reproducibility and to clarify what the aggregate figures do and do not capture.

\textit{Time window.} The reporting window spans five calendar months, from November~1, 2025 through March~31, 2026. All artifacts with a first-commit, first-submission, or first-deployment timestamp within this window are included; artifacts initiated earlier and merely touched during the window are excluded.

\textit{Data sources.} Software artifacts are enumerated from Git commit history across the research monorepo and satellite repositories (paper-mandate, paper-lattice, paper-trace, paper-context-engineering, paper-closed-loop, paper-augment-engineering, swiftspace-ai). Pipeline runs and lane transitions are pulled from the Jira REST API for the seven-lane workflow. Web deployments are counted from the Vercel deployments API for the swiftspace.ai project. Academic artifacts are counted from \texttt{.tex} source files and journal and preprint submission records. Curriculum artifacts are counted from the CE Curriculum Framework repository.

\textit{Counting rules.} For code, one artifact is defined as one source file at first merge into \texttt{main}; line counts are \texttt{cloc}-derived (comments and blanks excluded from the 41{,}392 production-code figure, included in the 84{,}294 total-with-tests figure). For papers, one artifact is one distinct manuscript advanced to at least draft-ready state within the window. For proposals, one artifact is one contracting-authority-submitted deliverable (G1--G5). For curriculum, one artifact is one lab block, one instructor support file, or one contact-hour-equivalent module. For web deployments, one artifact is one successful Vercel production build. Partial drafts, abandoned branches, and scratch experiments are not counted.

\textit{Domains with incomplete instrumentation.} Two domains could not be fully instrumented within the study window. HeyGen (video production) and Gamma.app (presentation design) expose only folder-level metadata through their MCP interfaces at the time of writing; per-artifact creation timestamps, iteration logs, and render counts were not retrievable via API. For these two domains, the paper reports qualitative evidence (domain entry without prior expertise, transfer velocity to first professional-quality output, stakeholder acceptance) but excludes them from the numerical aggregates. The Wright's Law curve in \Cref{fig:wrights-law} is therefore fitted to 82 instrumentally tracked artifacts; including the 10 excluded HeyGen/Gamma artifacts (reconstructed from folder-level metadata) would strengthen the production acceleration finding, so the 82-artifact fit represents a conservative lower bound.

\textit{Three datasets, three purposes.} This paper draws on three distinct datasets, each constructed for a different analytical purpose. (1)~The \textbf{200-interaction set} (102 Claude interactions plus 98 ChatGPT interactions) is the prompt-sophistication corpus. Each interaction was coded for prompt level (L1--L4), first-pass acceptance, and context-file count. This set drives \Cref{fig:prompt-quality} and the Cochran-Armitage trend test in \Cref{sec:cs-longitudinal}. (2)~The \textbf{102-interaction set} is the Claude-only subset of the 200-interaction set, used for the longitudinal phase analysis in \Cref{sec:cs-longitudinal} because only Claude interactions carry the phase metadata (tool sequence, handoff counts, context-package evolution) required for that analysis. (3)~The \textbf{82-artifact set} counts distinct work products (source files, manuscripts, deployments, lab blocks) rather than interactions. Each artifact maps to one or more interactions but the unit of analysis is the deliverable, not the conversation turn. This set drives the Wright's Law fit in \Cref{fig:wrights-law}. The 82 artifacts are instrumentally tracked; 10 additional HeyGen/Gamma artifacts are excluded for lack of per-artifact timestamps (\Cref{sec:cs-counting}).

\textit{Dataset overlap disclosure.} The 200-interaction corpus used for the prompt-sophistication analysis in \Cref{fig:prompt-quality} and the Cochran-Armitage test is the same corpus reported in the Context Engineering companion work~\cite{calboreanu2026context}. This paper re-analyzes it for the prompt-sophistication $\times$ first-pass-acceptance relationship; the Context Engineering paper analyzed it for structured-input sufficiency. No additional interactions were coded for the present paper.

\textit{Interaction-log provenance.} Claude and ChatGPT interaction logs were collected from each platform's native conversation-export feature (Claude Projects export and ChatGPT conversation history export). Interactions were harmonized into a single coding sheet recording: timestamp, platform (Claude/ChatGPT), turn boundaries, assigned prompt-sophistication level (L1--L4), first-pass acceptance (binary), iteration count to acceptance, and context-file count. Prompt-sophistication levels were assigned by the single practitioner-coder using the L1--L4 rubric defined in \ref{app:rubric}; inter-rater reliability was not assessed, which is acknowledged as a limitation.

\subsection{Tool Portfolio Inventory}\label{sec:cs-portfolio}

\begin{table*}[t]
\caption{Orchestration stack inventory for the case study practitioner: five \textbf{AI tools}, where prompt and context engineering skills are the primary mode of operation, and five \textbf{infrastructure components}, whose adoption follows traditional learning curves but which still participate in cross-tool workflows.}
\label{tab:tool-portfolio}
\centering
\small
\resizebox{\textwidth}{!}{%
\begin{tabular}{@{}llp{3.0cm}p{3.8cm}p{2.0cm}p{3.3cm}@{}}
\toprule
\textbf{Tool} & \textbf{Category} & \textbf{Domain(s)} & \textbf{PE/CE Skills Applied} & \textbf{Transfer Velocity} & \textbf{Key Work Products} \\
\midrule
Claude (Anthropic)
  & AI tool
  & Software dev., academic writing, proposals, curriculum, data analysis
  & Full PE + CE: authority docs, structured prompts, rubric validation, iterative refinement
  & Primary tool (baseline)
  & 41,392 LOC Python, 6 papers, 5 contract deliverables \\
Gamma.app
  & AI tool
  & Presentation design
  & PE: structured briefs as input; CE: brief documents as authority inputs
  & $\sim$2 hours
  & Briefing decks, training presentations$^{\dagger}$ \\
HeyGen
  & AI tool
  & Video production
  & PE: structured scripts; CE: scripts as authority documents
  & $\sim$4 hours
  & Training videos, demonstration content$^{\dagger}$ \\
MLX (Apple)
  & AI tool
  & ML fine-tuning
  & CE: training corpus as structured input
  & $\sim$1 week
  & 6 LoRA adapters \\
Tesseract OCR
  & AI tool
  & Document processing
  & CE: quality scoring pipeline as structured evaluation
  & $\sim$2 days
  & Document ingestion pipeline \\
\midrule
Jira (automated)
  & Infrastructure
  & Project management
  & Partial: pipeline config as structured workflow; automation required traditional software engineering$^{*}$
  & $\sim$1 week$^{*}$
  & 7-lane pipeline, 2,186+ tracked items \\
GitHub / Git
  & Infrastructure
  & Version control, CI/CD
  & CE: repository structure as authority; automated quality gates
  & Prior experience
  & 7+ repositories, automated audit rotation \\
Vercel
  & Infrastructure
  & Web deployment
  & PE + CE: deployment specs as structured inputs
  & $\sim$3 days
  & swiftspace.ai (production) \\
LaTeX / BibTeX
  & Infrastructure
  & Academic typesetting
  & CE: template structures, bibliography management
  & Prior experience
  & 7 paper repositories \\
WeasyPrint
  & Infrastructure
  & Data visualization
  & CE: dashboard templates as authority documents
  & $\sim$1 day
  & Weekly status dashboards \\
\bottomrule
\end{tabular}%
}\\[2pt]
{\scriptsize $^{*}$Jira automation required traditional Python scripting, not prompt/context engineering transfer; a boundary case discussed in \Cref{sec:cs-failures}.\\
$^{\dagger}$Per-artifact timestamps could not be retrieved via API; these domains are reported qualitatively and excluded from numerical aggregates (\Cref{sec:cs-counting}), and their transfer-velocity values are practitioner session-log estimates.}
\end{table*}

\Cref{tab:tool-portfolio} presents the complete orchestration-stack inventory. The portfolio comprises ten components in the orchestration stack: five AI tools (Claude, Gamma.app, HeyGen, MLX, Tesseract OCR) and five supporting infrastructure components (Jira, GitHub/Git, Vercel, LaTeX/BibTeX, WeasyPrint). Claude (Anthropic) served software development, technical and academic writing, proposal drafting, code review, data analysis, curriculum design, and systems architecture; Gamma.app served presentation design and visual communication; HeyGen served video production; Jira with integrated automation served project management and workflow orchestration; GitHub and Git served version control, collaborative development, and CI/CD; Vercel served web deployment and frontend engineering; MLX (Apple) served machine learning fine-tuning; Tesseract OCR served document processing; and WeasyPrint with Matplotlib served data visualization and report generation. For each tool, the practitioner documented which prompt and context engineering techniques transferred from prior tool experience, the time required to reach productive output (transfer velocity), the volume and quality of work products, and the validation methodology employed.

\subsection{Domain-by-Domain Analysis}\label{sec:cs-domains}

\Cref{tab:domain-outputs} summarizes outputs and validation signals by domain; the paragraphs below provide narrative detail.

The software development domain demonstrated the primary and most comprehensive application of augment engineering methodology. Operating within a governance-first research monorepo that integrates the specification layer~\cite{calboreanu2026mandate}, authorization layer~\cite{calboreanu2026lattice}, and execution and evidence layer~\cite{calboreanu2026trace} developed in this author's prior work into a single seven-lane automated pipeline, the practitioner generated 41,392 lines of production-quality Python source code (84,294 lines including test coverage). The accompanying prompt specifications comprised 7,914 lines of structured documentation across 68 distinct files, functioning as versioned context packages. Quality assurance yielded 781 unique test functions comprising 1,045 parameterized test cases; on the final commit of the study window, all 1,045 cases passed (100\%). The system processed 795 continuous pipeline runs across the seven automated lanes, with a dedicated 152-run evaluation phase achieving 100\% terminal success. Seven timestamped audit runs, conducted on 2026-02-17, all passed without exception. Transfer velocity in this domain was primary, as the prompt and context engineering methodologies were developed and refined here before application elsewhere.

Academic publication activities used Claude in combination with a standard academic tool chain. The practitioner advanced six papers into peer-review pipelines during the window; outcomes were mixed, spanning major-revision decisions, desk rejections, and papers still under review, with no final acceptances as of study close. Where major revisions were invited, revised versions were resubmitted. Context engineering proved transferable from software development workflow patterns directly: authority documents (research methodologies, standards frameworks), source data (evaluation corpora, empirical findings), and output constraints (journal templates, citation standards) were packaged using identical structural principles. Transfer velocity to first publishable draft was approximately one week.

Contract proposal and deliverable work produced five major contracting documents accepted by contracting authorities (designated G1 through G5 for operational security; specific program names withheld). The largest deliverable contained 469 paragraphs of technical prose, submitted through six sequential audit passes. Both prompt engineering (structured iteration with audit rubrics) and context engineering (compliance matrices as authority documents) transferred directly from software development. All deliverables achieved contracting authority acceptance. Transfer velocity to first draft was approximately two days.

Curriculum design activities produced comprehensive instructional materials without prior domain expertise. The practitioner designed 12 laboratory exercise blocks with facilitation guides and 96 instructor support files. Context engineering principles, particularly the methodology for assembling authority documents and output constraints, applied directly from software development. The curriculum framework document itself functioned as the master authority document, with Instructional System Design (ISD) and ADDIE methodologies as structural constraints. Transfer velocity was approximately one week from concept to structured instructional framework.

Video production via HeyGen represented the first domain in which the practitioner had zero prior experience. Training videos, stakeholder briefings, and demonstration content were produced with a transfer velocity of approximately four hours to first professional-quality output, estimated from the practitioner's session logs rather than artifact-level timestamps (see below). Prompt engineering techniques, particularly structured input formatting and iterative refinement, transferred effectively. Context engineering was adapted: scripts served as authority documents, with production requirements as constraints. Outputs were validated through use in stakeholder briefings, with client acceptance confirming quality. Unfortunately, per-artifact creation timestamps and iteration logs could not be retrieved from HeyGen through its MCP interface, which exposes folder-level metadata only. This domain is therefore reported qualitatively and is excluded from the numerical aggregates described in \Cref{sec:cs-counting}.

Presentation design through Gamma.app required zero graphic design expertise. Professional briefing materials, training presentations, and client deliverables were produced with a transfer velocity of approximately two hours to first professional-quality output, estimated from the practitioner's session logs rather than artifact-level timestamps (see below). Prompt engineering (structured inputs, iterative refinement) transferred directly; context engineering (brief documents as authority inputs) was adapted for visual communication. Outputs achieved client acceptance in stakeholder presentations. As with HeyGen, Gamma's MCP interface provides only folder-level read access, so deck-level creation dates and iteration counts could not be instrumented; Gamma is reported qualitatively and excluded from the aggregate counts (\Cref{sec:cs-counting}).

Web deployment demonstrated transfer of context and prompt engineering to production infrastructure. The practitioner deployed swiftspace.ai, a Next.js production application, on Vercel through a continuous-deployment workflow, with builds co-authored by Claude. Transfer velocity from project initiation to first production deployment was approximately three days, with four minutes elapsed from initial commit to live availability. The practitioner possessed minimal prior web development experience, yet context engineering (structured technical specifications) and prompt engineering transferred directly from software development domain work.

\begin{table*}[t]
\caption{Output summary by professional domain. Domains marked $\dagger$ are those where the practitioner had no specialist training prior to the study period. Volumes reported as ``n/a (qualitative)'' could not be retrieved via API instrumentation (\Cref{sec:cs-counting}) and are excluded from numerical aggregates. Validator independence varies across the quality signals; see \Cref{sec:disc-threats}.}
\label{tab:domain-outputs}
\centering
\small
\resizebox{\textwidth}{!}{%
\begin{tabular}{@{}lllp{3.5cm}p{3cm}@{}}
\toprule
\textbf{Domain} & \textbf{Tool(s)} & \textbf{Specialist?} & \textbf{Output Summary} & \textbf{Quality Validation} \\
\midrule
Software Development
  & Claude, GitHub, Jira
  & Yes (20+ yr)
  & 41,392 LOC Python source (84,294 incl.\ tests); 795+ pipeline runs; 781 test functions (1,045 parameterized); 152-run evaluation (100\% terminal success)
  & Automated test suite (100\% pass); 7 timestamped audit runs (all passing) \\
\addlinespace
Academic Publication
  & Claude + academic tool chain
  & No$\dagger$
  & 6 papers in peer review; 3,634 lines of structured source across 63 files
  & Advanced into peer review; mixed outcomes (major-revision, desk rejection, in-review); revised versions resubmitted where invited; 0 acceptances as of study close \\
\addlinespace
Proposals
  & Claude, Jira
  & Yes (contract)
  & 5 major deliverables; 469 paragraphs in largest (6 audit passes)
  & Contracting authority acceptance; multi-round audit compliance \\
\addlinespace
Curriculum Design
  & Claude
  & No$\dagger$
  & 12 lab blocks, 96 instructor support files
  & Subject-matter expert review; instructional design validation \\
\addlinespace
Video Production
  & HeyGen
  & No$\dagger$
  & Training and demonstration videos; n/a (qualitative, API-limited)
  & Stakeholder briefing acceptance \\
\addlinespace
Presentation Design
  & Gamma.app
  & No$\dagger$
  & Briefing decks, training presentations; n/a (qualitative, API-limited)
  & Stakeholder presentation acceptance \\
\addlinespace
Web Deployment
  & Claude, Vercel
  & No$\dagger$
  & swiftspace.ai (Next.js production app)
  & Production deployment on Vercel \\
\bottomrule
\end{tabular}%
}
\end{table*}

\subsection{Cross-Domain Workflow Examples}\label{sec:cs-workflows}

The four case-study pipelines illustrate representative multi-tool orchestration patterns. The academic paper pipeline exemplifies authority-document-driven composition: a companion paper on specification-layer governance~\cite{calboreanu2026mandate} was drafted in Claude using a context package containing the NIST AI Risk Management Framework~\cite{nist-ai-rmf} as authoritative reference, the research monorepo's evaluation corpus as empirical source material, and a Springer journal template as output constraint. The paper was then typeset and version-controlled through a standard academic tool chain, with rubric-based review at each governance checkpoint before final submission.

The contract deliverable pipeline demonstrates prompt engineering applied through governance checkpoints. A client technical evaluation response (G2) was drafted in Claude using a context package containing the RFI requirements matrix as authority, client formatting standards as constraints, and prior successful proposals as exemplars. The resulting 469-paragraph response underwent six sequential audit passes, each tracked in Jira and linked to the governance framework. Quality gates included compliance review after each Claude draft iteration, section-level audit against RFI requirement matrices, and final red-team review before submission to contracting authority. Acceptance by the contracting authority confirmed the methodology's effectiveness for compliance-driven composition.

The training content pipeline demonstrates nested context engineering across three tools. A prompt engineering curriculum was designed in Claude using an instructional design framework as master authority document and ADDIE/ISD methodology as structural constraint. Output, comprising 12 lab blocks with facilitation guides and 96 instructor support files, was structured for downstream tool consumption. Lab content was repackaged as Gamma briefs to drive training presentation production. Selected lab walkthroughs were scripted for HeyGen video production. Quality gates included subject-matter expert review of curriculum content, visual QA of Gamma slides, and content validation of HeyGen videos. HeyGen functions as a terminal node in this workflow; video outputs cannot serve as structured inputs to other tools and require separate hosting and embedding, limiting orchestration flexibility.

The product development pipeline demonstrates fully automated integration through governance infrastructure, following the closed-loop Jira-integrated orchestration pattern described in companion work~\cite{calboreanu2026closedloop}. The research monorepo was developed through continuous pipeline runs across seven automated Jira lanes, with Claude generating code from versioned prompt specifications as context packages. Each lane transition served as an automated quality gate covering specification review, implementation, testing, audit, integration, documentation, and deployment, with the production application deployed through standard CI/CD infrastructure. This workflow demonstrates the highest level of orchestration integration, with minimal manual handoff overhead.

\subsection{Skill Portability Evidence}\label{sec:cs-portability}

\begin{table*}[t]
\caption{Skill portability evidence across the tool portfolio. For each tool, the table
assesses the four transfer dimensions defined in Phase~3 of the methodology: input
structuring (can the tool accept structured inputs?), output validation (can rubric-based
evaluation be applied?), iteration patterns (does generate-evaluate-revise transfer?),
and integration compatibility (can outputs serve as inputs to other tools?). Ratings:
\textbf{D}~=~direct transfer, \textbf{A}~=~adapted transfer, \textbf{N}~=~not applicable.}
\label{tab:skill-portability}
\centering
\small
\resizebox{\textwidth}{!}{%
\begin{tabular}{@{}lccccp{5.5cm}@{}}
\toprule
\textbf{Tool} & \textbf{Input Struct.} & \textbf{Output Valid.} & \textbf{Iteration} & \textbf{Integration} & \textbf{Notes} \\
\midrule
Claude       & D & D & D & D & Primary tool; all PE/CE skills apply directly \\
Gamma.app    & A & D & D & A & Brief documents adapt as authority inputs; export requires formatting \\
HeyGen       & A & D & A & N & Scripts serve as structured input; iteration limited by rendering time; video outputs cannot serve as structured inputs to other tools (requires separate hosting/embedding), making HeyGen a terminal node in cross-domain workflows \\
Jira         & A & A & A & D & Automation config as structured input; validation via workflow metrics \\
GitHub/Git   & D & D & D & D & Repository structure and CI as context engineering artifacts \\
Vercel       & A & D & A & D & Deployment config as structured input; iteration via preview deployments \\
LaTeX/BibTeX & D & D & D & D & Template structure and bibliography as context packages \\
MLX          & A & A & A & N & Training corpus as structured input; evaluation via loss metrics \\
Tesseract    & A & D & A & D & Pipeline config as structured input; quality scoring transfers \\
WeasyPrint   & D & D & D & D & Templates as authority documents; output directly distributable \\
\bottomrule
\end{tabular}%
}
\end{table*}

\begin{table*}[t]
\caption{Portability metric measurements across four representative cross-domain
workflows. Transfer velocity ($T_v$) is time from first tool use to professional-quality
output. Cross-domain output quality ($Q_d$) is validated through external signals.
Orchestration overhead ($O_h$) is estimated as the proportion of total workflow time
spent on cross-tool coordination (formatting handoffs, managing quality gates, resolving
integration failures) versus in-tool production. Coverage breadth ($C_b$) is the number
of distinct professional domains served by the workflow.}
\label{tab:portability-metrics}
\centering
\small
\resizebox{\textwidth}{!}{%
\begin{tabular}{@{}lp{2cm}p{2.5cm}p{2cm}cp{3cm}@{}}
\toprule
\textbf{Workflow} & \textbf{Tools} & \textbf{$T_v$ (first productive output)} & \textbf{$O_h$ (coordination \%)} & \textbf{$C_b$} & \textbf{$Q_d$ validation} \\
\midrule
Academic Paper Pipeline
  & Claude + academic tool chain
  & $\sim$1 week to first publishable draft
  & $\sim$12\%
  & 2
  & Mixed peer-review outcomes (major-revision, desk rejection, in-review); 2 major-revision decisions received; 0 acceptances as of study close \\
\addlinespace
Training Content Pipeline
  & Claude, Gamma, HeyGen
  & $\sim$4 hours (HeyGen); $\sim$2 hours (Gamma)
  & $\sim$18\%
  & 3
  & Instructional design validation; stakeholder briefing acceptance \\
\addlinespace
Product Development Pipeline
  & Claude, Jira, GitHub, Vercel
  & $\sim$3 days (Vercel); 4 min to first deploy
  & $\sim$8\%
  & 2
  & Production deployment; 100\% test pass; continuous Vercel deployment, Claude co-authored \\
\addlinespace
Contract Deliverable Pipeline
  & Claude, Jira
  & $\sim$2 days to first deliverable draft
  & $\sim$5\%
  & 1
  & Contracting authority acceptance; all 5 submissions accepted \\
\bottomrule
\end{tabular}%
}

\medskip
\noindent\small\textit{Note:} $O_h$ estimates are derived from practitioner time logs
and workflow timestamps (Git commits, Jira ticket transitions, Vercel deployment records).
The Training Content Pipeline has the highest overhead because HeyGen is a terminal
node: video outputs require separate hosting and cannot serve as structured inputs to
downstream tools, creating manual integration steps. The Product Development Pipeline
has the lowest overhead because Jira-as-Code automates cross-tool authority transfer
and GitHub CI provides automated quality gates, reducing manual coordination to
near-zero for routine deployments. Per-pipeline $C_b$ values overlap across pipelines; the practitioner-level $C_b = 7$ is their union, not their sum.
\end{table*}

\Cref{tab:skill-portability} documents which prompt and context engineering techniques transferred to which tools and which required domain-specific adaptation. Four key findings emerged. First, transfer velocity showed a directional decrease at the phase level (Phase~1--2 adoptions required days to weeks, Phase~3 adoptions hours to a few days) consistent with cumulative competency development, though per-tool values did not decrease monotonically with adoption order (see the curriculum and Vercel exceptions discussed below). Second, input structuring, the core context engineering capability, transferred most directly across all tools. The methodology for assembling a context package for Claude was architecturally identical to the methodology for structuring a Gamma brief or HeyGen script: identify authority documents, gather source material, establish output constraints, format for tool consumption, and validate against rubrics. Third, output validation (prompt engineering methodology) transferred without modification across tools. The rubric-based evaluation approach developed for Claude applied directly to Gamma presentations and HeyGen videos. Fourth, integration compatibility varied substantially depending on tool output format. Claude's markdown-formatted output integrated easily with LaTeX and other structured formats. GitHub's version-controlled files integrated seamlessly with CI/CD infrastructure. HeyGen videos, by contrast, required manual hosting and embedding decisions, making cross-tool orchestration more difficult.

\Cref{tab:portability-metrics} estimates integration overhead ($O_h$) across four representative workflows. Overhead values are practitioner estimates reconstructed post-hoc from Jira lane transition logs, Git commit cadence, and time allocated to cross-tool format translation; continuous sub-task time tracking was not instrumented during the study window, so these values should be treated as approximate ratios rather than precisely measured proportions. The academic paper pipeline incurred an estimated 12\% overhead (academic typesetting and reference validation). The training content pipeline incurred approximately 18\% overhead, attributable primarily to HeyGen's terminal-node status requiring manual integration. The product development pipeline, leveraging Jira-as-Code automation for handoff management, incurred an estimated 8\% overhead. The contract deliverable pipeline, with minimal cross-tool coordination and single-pass integration, incurred roughly 5\% overhead. These directional findings are consistent with the decomposition in \Cref{eq:oh-decomp}: $O_h$ is lowest where Jira-as-Code automates authority transfer between tools and highest where terminal nodes (particularly HeyGen video) require manual integration steps. Precise measurement of $O_h$ through continuous time tracking is identified as a future-work instrumentation target (\Cref{sec:disc-future}).

The empirical transfer velocity data are directionally consistent with the power-law model of \Cref{eq:tv} but require careful interpretation. Six non-primary tool adoptions were recorded (one per domain entered, not per tool): academic publication ($\sim$168 hrs), contract proposals ($\sim$48 hrs), curriculum design ($\sim$168 hrs), HeyGen video ($\sim$4 hrs), Gamma presentations ($\sim$2 hrs), and Vercel web deployment ($\sim$72 hrs). Importantly, $T_v$ does not decrease monotonically with adoption order: curriculum design required $\sim$168 hours despite being the third cross-domain tool, and Vercel required $\sim$72 hours despite being the sixth. These deviations reflect domain-specific factors: curriculum design was the first domain requiring pedagogical structure (not merely prompt transfer), and Vercel involved traditional web development alongside prompt engineering.

At the phase level, however, an apparent trend is visible: the three Phase~1--2 adoptions ranged from 48 to 168 hours (contract proposals at the low end, academic publishing and curriculum design at the high end), while the three Phase~3 adoptions ranged from 2 to 72 hours (HeyGen and Gamma at a few hours each, Vercel a 72-hour outlier driven by traditional web-development work rather than meta-skill transfer). The six-tool sequence is directionally consistent with the power law in \Cref{eq:tv} ($\hat{\beta} \approx 1.57$, $R^2 = 0.29$, $p > 0.10$) but not statistically powered: with only six post-primary observations the confidence interval spans zero. The Cochran-Armitage trend test ($n = 200$, $p < 0.01$; \Cref{sec:cs-longitudinal}) and the Wright's-Law cross-artifact analysis ($n = 82$, $p < 0.01$; \Cref{fig:wrights-law}) are the primary statistical evidence for the prompt-quality association and portfolio-wide production acceleration that the framework predicts; neither test directly measures cross-tool transfer. Multi-practitioner studies with larger tool samples are required to estimate $\beta$ at the tool level and are identified as the highest-priority replication target (\Cref{sec:disc-future}).

Decomposing $O_h$ across the four pipelines reveals three structural components:
\begin{equation}\label{eq:oh-decomp}
  O_h \approx O_{\text{floor}} + O_{\text{terminal}} - O_{\text{auto}}
\end{equation}
where $O_{\text{floor}} \approx 5\%$ represents the irreducible coordination overhead present in even the simplest pipeline (the contract deliverable baseline), $O_{\text{terminal}}$ is the penalty incurred per terminal node whose outputs cannot serve as structured inputs to downstream tools, and $O_{\text{auto}}$ is the automation dividend from Jira-as-Code or CI/CD-mediated handoffs. This decomposition is approximate; residual overhead from pipeline complexity (number of tool boundaries, format heterogeneity) is not captured by the three-term model. The contract deliverable pipeline ($O_h \approx 5\%$) represents the floor with minimal cross-tool coordination. The training pipeline ($O_h \approx 18\%$) pays the largest penalty, attributable to HeyGen's terminal-node status ($O_{\text{terminal}} \approx 13\text{pp}$ above floor). The academic pipeline ($O_h \approx 12\%$) incurs moderate overhead from academic typesetting across multiple tool boundaries ($\sim$7pp above floor). The product pipeline ($O_h \approx 8\%$) benefits from Jira-as-Code automation but still incurs 3pp above floor from the sheer number of tool transitions in a seven-lane pipeline. The consistent finding across all four pipelines is that terminal nodes and format translation at tool boundaries are the primary overhead drivers, and automation at high-frequency boundaries reduces but does not eliminate coordination cost.

\subsection{Failure Cases}\label{sec:cs-failures}

Not all augmentation attempts succeeded, and these failure cases define the boundaries of augment engineering methodology. Hardware procurement and physical lab infrastructure setup required physical actions outside any AI tool's applicability; augment engineering does not claim to augment physical work. Personnel management, contract negotiation, and administrative authorization require human judgment, relationship management, and legal authority; AI tools could provide drafting support but could not replace human decision-making. Classified briefing preparation encountered security constraints that limited AI tool use to unclassified content; classified portions remained fully human-produced. Fine-grained visual design sometimes exceeded Gamma's capabilities: highly specific visual requirements (exact pixel positioning, brand-compliant complex layouts) occasionally necessitated manual refinement, though prompt and context engineering reduced but did not eliminate iteration cycles.

Two failures deserve special emphasis. Jira pipeline automation required traditional software engineering (Python scripting, REST API integration, and systems configuration) rather than prompt and context engineering skill transfer. The ${\sim}$1\,week ``transfer velocity'' for Jira reflects conventional software development time, not meta-skill portability. This defines an important boundary: tools requiring programmatic configuration rather than structured natural-language input are outside the augment engineering framework. They can participate in orchestration portfolios, but their adoption follows traditional learning curves, not prompt/context engineering transfer. This boundary motivates the AI-tool / infrastructure split in \Cref{tab:tool-portfolio}. Components classified as infrastructure participate in the orchestration stack but are not within the scope of the portability claim, which concerns prompt and context engineering skill transfer across AI tools specifically. HeyGen integration compatibility similarly defines a boundary condition. Video outputs are terminal artifacts that cannot serve as structured inputs to downstream tools. Unlike Claude markdown (easily consumed by downstream typesetting and presentation tools) or GitHub-managed files (consumed by CI/CD), HeyGen videos require separate hosting infrastructure and manual embedding. This makes HeyGen a terminal node in orchestration chains, not an intermediate tool, limiting workflow flexibility. Integration design (which the framework defines as the critical orchestration phase) must account for such terminal nodes and position them at workflow ends rather than midpoints.

These failures prevent overclaiming and establish boundary conditions: augment engineering is effective for tasks that (a) can be mediated by AI tools accepting structured inputs, (b) accept prompt and context engineering skill application, and (c) produce outputs compatible with downstream tools in the orchestration chain. When any condition fails, the practitioner reverts to traditional methods for that specific tool or domain segment, sustaining the overall work through hybrid augmented-traditional approaches.

\subsection{Longitudinal Observations}\label{sec:cs-longitudinal}

\begin{figure}[t]
  \centering
  \includegraphics[width=0.85\textwidth]{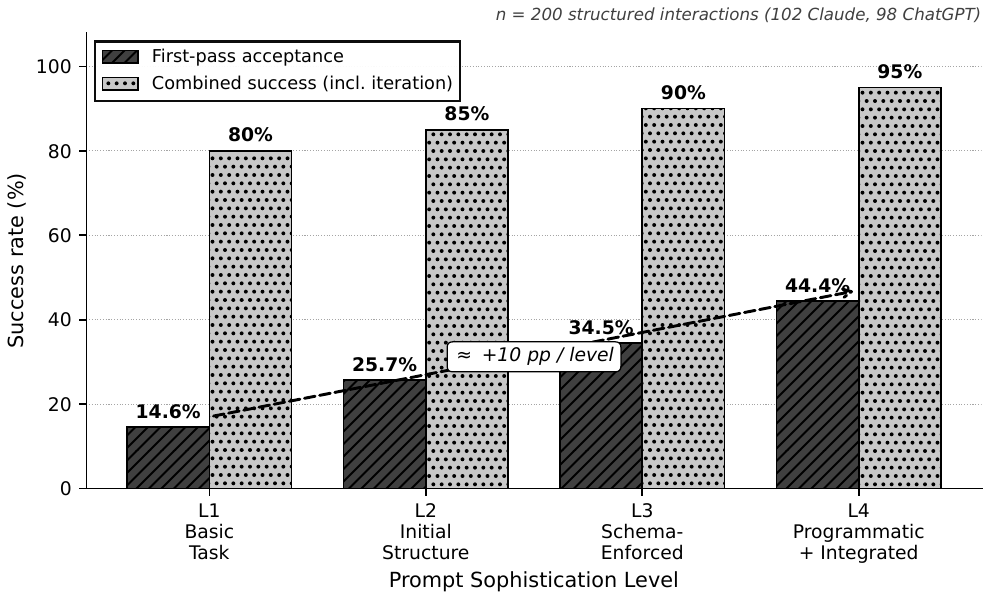}
  \caption{Prompt sophistication level versus output quality across 200
  structured interactions (102 Claude, 98 ChatGPT). First-pass acceptance
  rises by approximately 10 percentage points per level: L1 14.6\% (7/48),
  L2 25.7\% (18/70), L3 34.5\% (19/55), L4 44.4\% (12/27). Combined success
  including iteration ranges from roughly 80\% (L1) to 95\% (L4). The
  residual L4 failures were infrastructure-bound, arising from the OpenAI
  Codex execution environment used for certain interactions rather than
  from prompt quality. This gradient supports the narrower claim that more structured prompting is associated with higher first-pass acceptance within these two chat LLMs. It does not test \Cref{eq:tv}, which models cross-tool transfer velocity, nor does it test portfolio-wide portability.}
  \label{fig:prompt-quality}
\end{figure}

\begin{figure*}[t]
  \centering
  \includegraphics[width=\textwidth]{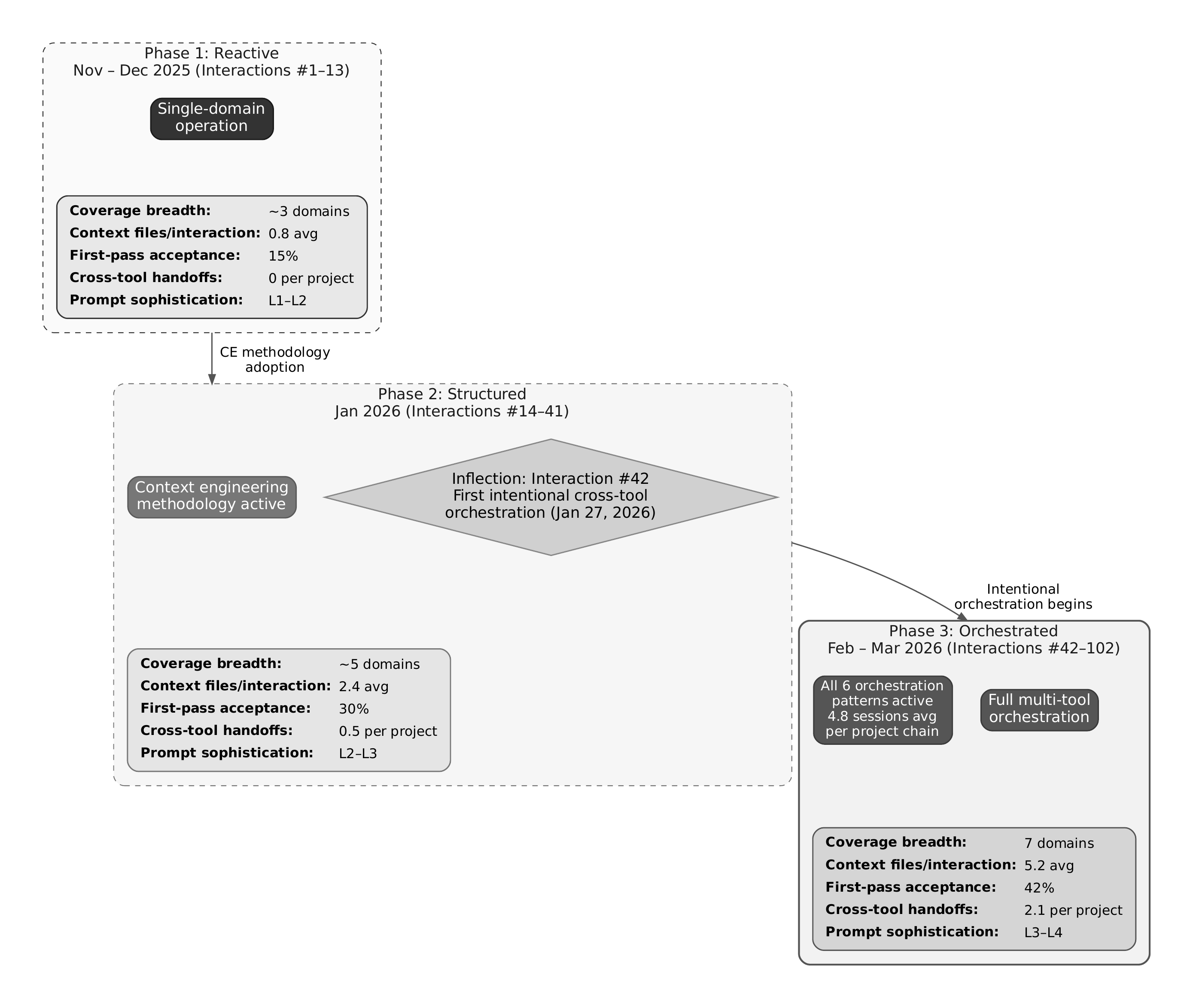}
  \caption{Longitudinal capability evolution across three phases of the
  five-month study. Coverage breadth expanded from 3 to 7 domains,
  first-pass acceptance rose from 15\% to 42\%, and cross-tool handoffs
  increased from 0 to 2.1 per project. The inflection point at interaction
  \#42 (January 27, 2026) marks the first intentional cross-tool
  orchestration.}
  \label{fig:longitudinal}
\end{figure*}

\begin{figure*}[t]
\centering
\includegraphics[width=\textwidth]{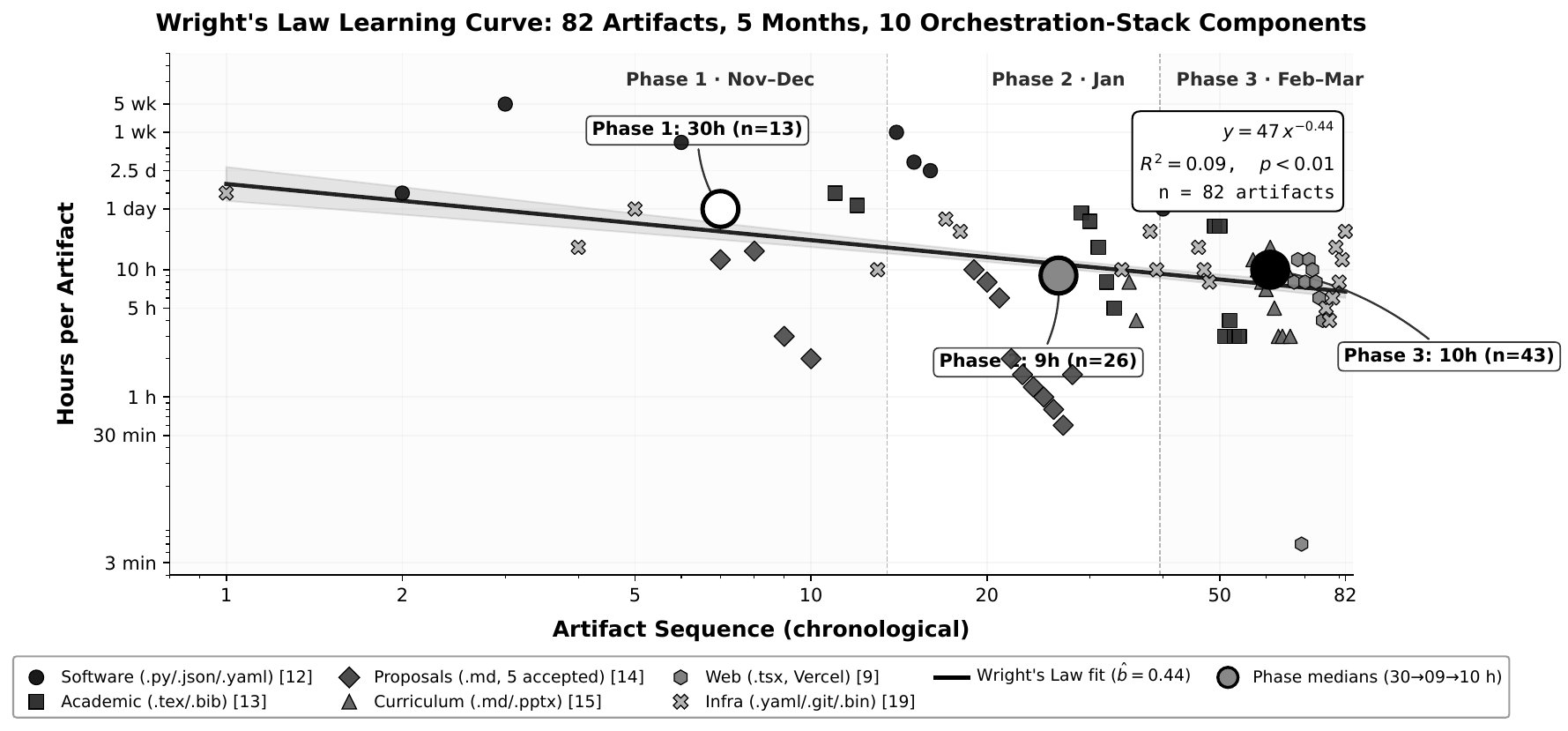}
\caption{Wright's Law learning curve across 82 instrumentally tracked sub-deliverables produced
by a single practitioner over five months using ten orchestration-stack components across seven
professional domains. Ten HeyGen/Gamma artifacts are excluded because per-artifact timestamps could not be retrieved via API (\Cref{sec:cs-counting}). Each point represents one artifact; marker shape
and grayscale shade jointly denote the six instrumented domain categories
(legend, bottom). An ordinary-least-squares fit on the log-transformed
data yields the power-law model $y = 47 \cdot x^{-0.44}$
($R^2 = 0.09$, slope $t = -2.85$, $p < 0.01$, $n = 82$). The slope is
statistically significant at the $\alpha = 0.01$ level; the modest $R^2$
reflects substantial per-artifact variance in complexity (some artifacts
absorbed weeks of effort, others minutes), which the aggregate power law
intentionally does not control for. Phase-level medians (large
circled markers) show the learning trajectory: 30\,h/artifact in Phase~1
(tool onboarding, $n = 13$), 9\,h in Phase~2 (cross-domain integration, $n = 26$), and 10\,h in
Phase~3 (portfolio-scale orchestration, $n = 43$). A Kruskal-Wallis test
across the three phases does not reach significance at $\alpha = 0.05$
($H = 5.01$, $p = 0.08$); pairwise Mann-Whitney comparisons show moderate
effect sizes for Phase~1 vs.\ 2 ($r = -0.41$, $p = 0.04$) and Phase~1
vs.\ 3 ($r = -0.38$, $p = 0.04$), neither of which survives Bonferroni
correction ($\alpha_{\text{adj}} = 0.017$). Phases~2 and~3 are not
distinguishable ($p = 0.69$). The phase-level acceleration is therefore
directional, not statistically confirmed at conventional thresholds. Industry baselines for full-deliverable production (e.g., the Chapman Alliance curriculum-development ratios~\cite{chapman2010learning}) are not unit-matched to this paper's artifact unit, which spans source files, manuscripts, deployments, and lab blocks at differing granularity. Such baselines are not plotted here; cross-study comparison requires unit harmonization, identified as a future-work instrumentation target (\Cref{sec:disc-future}). The 95\% confidence
interval (shaded band) accounts for variance in artifact complexity across
domains. Note that this figure measures production acceleration across the full artifact portfolio, not specifically skill portability across tool boundaries; the latter is tested by the transfer-velocity model (\Cref{eq:tv}) with a smaller, underpowered sample.}
\label{fig:wrights-law}
\end{figure*}

Longitudinal analysis reveals accelerating capability development over the five-month study period, documented through detailed interaction logs and evidence map audits. Phase 1 (November--December 2025, interactions \#1--13) was characterized by reactive, single-domain work. The practitioner operated primarily in software development and contract proposals using Claude and Jira, with minimal cross-tool coordination. Context packages averaged 0.8 files per interaction, reflecting surface-level authority document integration. First-pass acceptance averaged approximately 15\%,\footnote{Phase-level first-pass acceptance rates are means over the Claude-only 102-interaction longitudinal subset (\Cref{sec:cs-counting}), partitioned by interaction index into Phase~1 ($n = 13$), Phase~2 ($n = 28$), and Phase~3 ($n = 61$).} indicating substantial iteration cycles. Zero cross-tool handoffs occurred; each project operated in isolation. Prompt sophistication ranged from L1 (basic task articulation) to L2 (initial structure). Coverage breadth encompassed two to three professional domains.

Phase 2 (January 2026, interactions \#14--41) marked the transition to structured context engineering methodology. Context packages evolved to 2.4 files per interaction with distinct functional roles (authority documents, source material, constraints). All four context engineering pipeline stages (authority, source, constraints, validation) became consistently active. First-pass acceptance rose to approximately 30\%, indicating improved prompt quality. The first cross-tool material passing appeared in this phase, manual and unstructured, at roughly 0.5 instances per project; intentional cross-tool orchestration, in which a handoff is deliberately designed with a structured intermediary, had not yet begun. Prompt sophistication advanced from L2 to L3 (schema-enforced structure). Coverage breadth expanded to approximately five domains. The boundary into Phase~3 is the critical inflection point at interaction \#42 (January 27, 2026), where the practitioner first \emph{designed} a cross-tool workflow: Claude generated an audit prompt specifically structured for ChatGPT consumption. This first intentional multi-tool integration is treated as the start of Phase~3.

Phase 3 (February--March 2026, interactions \#42--102) achieved full multi-tool orchestration capability. Context packages averaged 5.2 files per interaction, supporting complex nested workflows. First-pass acceptance reached approximately 42\%, reflecting mature prompt engineering discipline. Cross-tool handoffs averaged 2.1 per project, with projects chaining 4.8 sessions on average across tool boundaries. Prompt sophistication reached L3--L4 (programmatic, schema-enforced, and infrastructure-integrated). Coverage breadth encompassed all seven professional domains. Skill transfer accelerated over time: augment engineering competence itself proved to be a learnable capability, developing through deliberate practice and systematic methodology refinement rather than representing a fixed trait. The practitioner did not enter the study with multi-tool orchestration expertise; this competency emerged through application of portable prompt and context engineering principles to increasingly complex integration challenges. \Cref{fig:longitudinal} visualizes this three-phase evolution with metrics at each stage.

Across all 200 structured interactions (102 Claude plus 98 ChatGPT), first-pass acceptance rises with prompt sophistication level: L1 14.6\% (7/48), L2 25.7\% (18/70), L3 34.5\% (19/55), and L4 44.4\% (12/27), an increase of approximately 10 percentage points per level (\Cref{fig:prompt-quality}). Combined success including iteration ranges from roughly 80\% at L1 to 95\% at L4. These are descriptive category means for four ordered levels, not a regression fit to individual observations; the formal test of the trend is the Cochran-Armitage statistic reported next.

The association between prompt level and first-pass acceptance is statistically significant. A Cochran-Armitage trend test applied to the raw 200-interaction contingency table (L1: 7/48, L2: 18/70, L3: 19/55, L4: 12/27 first-pass successes) yields $Z = 3.04$, $p < 0.01$, indicating a monotone increasing trend in acceptance rate across ordered prompt sophistication levels. A chi-square test of independence yields $\chi^2 = 9.26$ ($\mathit{df} = 3$, $p < 0.05$) with Cram\'{e}r's $V = 0.22$, indicating a small-to-medium effect size. The trend is genuine; higher prompt sophistication is associated with higher first-pass acceptance across 200 interactions. This trend test provides the strongest statistical evidence in this paper for the within-tool prompt-quality association. It does not test cross-tool skill portability, which remains a directional finding awaiting multi-practitioner replication (\Cref{sec:disc-future}).

Iteration cost, the fraction of interactions requiring revision, falls as sophistication rises: at L4 roughly half of interactions require any iteration, compared with about 65\% at L1. The residual L4 failures were infrastructure-bound rather than quality-related, arising from constraints in the OpenAI Codex execution environment used for certain L4 coding interactions; Codex is an external execution environment, not part of the orchestration portfolio in \Cref{tab:tool-portfolio}. This suggests the true quality ceiling of L4 prompting may exceed the measured 95\%.

Template compounding provides additional evidence for skill transfer acceleration. Context engineering templates, structured input packages reused across multiple interactions, appear to exhibit a power-law cost reduction of the form $C(n) = C_0 \cdot n^{-\alpha}$, where $C_0$ is the initial creation cost, $n$ is the number of reuses, and $\alpha$ is the amortization exponent. The academic publication pipeline, for instance, required 120 minutes on first use and 45 minutes on third use, yielding an illustrative exponent of $\alpha \approx 0.89$ if a power-law form is assumed. This estimate is derived from two time observations; no confidence interval can be constructed, and $\alpha$ should not be compared directly with the $p$-valued statistics reported elsewhere. If the exponent is approximately correct, it implies aggressive cost reduction: by the sixth reuse, a template would cost approximately 20\% of its original creation time. Across the study, eight distinct template families were reused a total of 67 times, with the PE Curriculum Framework alone serving as a standing authority document across 12+ interactions. Template compounding is the mechanism through which context engineering skills amortize across the portfolio, reducing $O_h$ by eliminating repeated specification effort.

First-pass acceptance rose monotonically with context file count across phases (Phase~1: 0.8~files, ${\approx}15\%$; Phase~2: 2.4~files, ${\approx}30\%$; Phase~3: 5.2~files, ${\approx}42\%$), consistent with the context engineering thesis that structured input design is the primary lever for output quality; the functional form cannot be identified from three observations. These phase-level first-pass rates aggregate interactions by time period; they are distinct from the L1--L4 rates, which partition the same interactions by prompt sophistication. The two analyses are independent, and their numerical similarity reflects that later-phase interactions concentrate at higher prompt levels. Coverage breadth expanded by four professional domains over the five-month window (3 domains in Phase~1, 5 in Phase~2, 7 in Phase~3), a sustained domain-entry rate enabled by cumulative transfer velocity improvements. Cross-tool handoffs accelerated from zero per project in Phase~1 to 2.1 per project in Phase~3, with handoff frequency growing faster than domain count, reflecting increasing orchestration complexity as the practitioner transitioned from parallel single-tool operation to integrated multi-tool pipelines. Whether these trends plateau, accelerate, or remain linear over longer time horizons is an empirical question for multi-practitioner replication.

\section{Discussion}\label{sec:discussion}

This section addresses the scope and limitations of the augment engineering framework, situates it within existing empirical work on domain transfer and skill generalization, and proposes directions for future replication.

\subsection{Validity of the Single-Practitioner Foundation}\label{sec:disc-limitations}

The formative evidence presented in this paper rests on one practitioner at one site over a five-month period. This limitation is acknowledged explicitly. The case study is not positioned as statistically generalizable evidence of the framework's efficacy across populations; rather, it serves as foundational work establishing the descriptive accuracy and prescriptive utility of the methodology. Statistical generalizability would require multi-practitioner replication with randomized assignment, controlled task conditions, and standardized measurement protocols, all of which are proposed as future work. The single-practitioner design does permit detailed longitudinal observation of skill transfer across multiple domains and tools, which is precisely the goal of a formative case study. Broader replication is necessary before the framework can be deployed organizationally, but does not invalidate the framework's logical coherence or the observation that domain entry occurred in this instance.

A second confound complicates interpretation: the practitioner brought significant prior experience in several of the domains studied (software engineering, cybersecurity, and contract proposal writing). A skeptical reviewer might attribute cross-domain success to general cognitive ability or domain expertise rather than to the portability of prompt and context engineering skills. This objection is addressed in detail in \Cref{sec:disc-expertise}, but the core response deserves mention here: the practitioner had no specialist training in video production, presentation design, curriculum design, academic publishing, or web deployment. These domains were entered entirely via augment engineering methodology applied to purpose-built AI tools. The absence of formal expertise in video production, yet the production of professional-grade deliverables within hours of tool use, does not comport with the hypothesis that general intelligence explains the outcome. The expertise confound is real and significant for the domains where the practitioner had deep background (software development, cybersecurity, contract proposals); it does not explain domain entry in domains where no prior training existed.

A third limitation concerns output quality assessment. The paper relies partly on external quality signals (advancement past desk-review to substantive peer review, stakeholder sign-off, production deployment) and partly on practitioner judgment. Where possible, objective metrics (test pass rates, code audit convergence) are reported directly. The retrospective application of the six-phase framework is a fourth limitation: the practitioner did not prospectively follow the methodology but rather applied augment engineering intuitively before the framework was codified. This weakens causal claims about the methodology's necessity but does not invalidate the framework's descriptive fit to the observed behavior or its utility as a prescriptive guide for future practitioners.

\textit{Ethics statement.} This study involved a single practitioner analyzing their own work products and tool interactions. No human subjects other than the author-practitioner were involved, no personally identifiable information from third parties was collected, and no experimental interventions were applied to external participants. As a self-study of professional practice, this research falls outside the scope of institutional review board (IRB) oversight. All data were collected from the practitioner's own operational records (Git commits, Jira logs, pipeline metrics) under routine professional activity. Future multi-practitioner replication studies will require IRB review and informed consent protocols.

\subsection{Threats to Validity}\label{sec:disc-threats}

Beyond the scope limitations discussed above, six specific threats to validity warrant explicit treatment, organized by three categories of the Wohlin et al. taxonomy (construct, internal, and external validity)~\cite{wohlin2012experimentation}; conclusion-validity concerns are addressed throughout the paper as explicit statistical-power and sample-size caveats. \Cref{tab:safeguards} summarizes the independence level of each measurement class used in the paper, consolidating the three-class evidence framing introduced in \Cref{sec:cs-counting}.

\begin{table}[h]
\caption{Measurement classes and their independence level. Claims in this paper are scoped to the evidence class that produced them; see \Cref{sec:cs-counting} for per-claim attribution.}
\label{tab:safeguards}
\centering
\small
\begin{tabular}{@{}p{4.2cm}p{3.8cm}p{3.2cm}@{}}
\toprule
\textbf{Measurement} & \textbf{Method} & \textbf{Independence} \\
\midrule
Interaction counts, artifact volumes, phase assignments
  & Practitioner-maintained logs and manifest
  & Internal \\
\addlinespace
Git commits, Jira lane transitions, test pass rates (781 functions, 1{,}045 cases), audit-run convergence, deployment counts
  & Automated system records extracted post-hoc
  & Objective / instrumented \\
\addlinespace
Advancement past desk-review to substantive peer review, contracting authority acceptance, production deployment
  & External evaluators outside the practitioner's control
  & Independent (with caveats for validator relationships, \Cref{sec:disc-threats}) \\
\addlinespace
Prompt-sophistication level (L1--L4) and first-pass acceptance coding for 200 interactions
  & Single practitioner-coder applying the rubric in \ref{app:rubric}
  & Internal; no inter-rater reliability assessed \\
\bottomrule
\end{tabular}
\end{table}

\textit{Construct validity: self-observation and Hawthorne effects.} The practitioner is simultaneously the framework designer, the subject, and the data collector. Awareness of being observed, particularly awareness of which behaviors will constitute evidence for the framework, can shape those behaviors. This threat is partially mitigated by the fact that the operational records (Git commits, Jira lanes, Vercel deployments, audit logs) were generated as routine professional artifacts under pre-existing governance controls rather than created for the study, and that external quality signals (advancement to substantive peer review, contracting authority acceptance, production deployment) are not controllable by the practitioner. It cannot be fully eliminated in a self-study and should be assumed present.

\textit{Internal validity: selection bias in tool and artifact inclusion.} The tool portfolio was not sampled from a population of candidate tools; it reflects tools the practitioner elected to adopt. Artifacts excluded from the quantitative aggregates (notably HeyGen and Gamma, see \Cref{sec:cs-counting}) are excluded because of instrumentation limits, not outcome. Yet the set of tools that entered the portfolio in the first place is shaped by the practitioner's prior expectations about which tools would succeed, producing a survivorship bias in favor of tools with portable input models. The framework's central claim (that prompt and context engineering skills transfer) is therefore tested on tools pre-filtered for compatibility with that claim. Multi-practitioner replication with an assigned rather than self-selected portfolio is the direct remedy.

\textit{Internal validity: measurement bias in self-reported effort.} Transfer velocity values for several domains are reported in coarse units (``approximately four hours,'' ``approximately one week'') because continuous time tracking was not instrumented at the sub-hour level for domains outside the Jira-tracked software pipeline. For quantitatively precise claims (e.g., the Wright's Law fit), only instrumentally measured intervals are used. Readers should treat the coarse transfer-velocity figures as order-of-magnitude rather than precise measurements.

\textit{External validity: temporal generalizability.} The study spans November 2025 through March 2026, a period of rapid underlying model improvement. The frontier models advanced materially within the window. On the Claude side, Sonnet~4.5 (September~2025) and Haiku~4.5 (October~2025) were current at the study's start; Opus~4.5 was released in late November~2025, during Phase~1; and Opus~4.6 and Sonnet~4.6 were both released in February~2026, during Phase~3. Successive ChatGPT model updates continued in parallel throughout. Phase~1 (Nov--Dec) and Phase~2 (Jan) thus operated largely within a single model generation, while Phase~3 (Feb--Mar) coincided with two further frontier releases. Effects attributed to practitioner learning may therefore be partly attributable to exogenous tool improvement, with the largest exposure in Phase~3; the study design cannot decompose these two sources, as discussed in \Cref{sec:disc-future}. Additionally, the tool landscape itself may evolve such that specific portfolio choices reported here are obsolete by the time of publication; the framework-level contributions (six-phase methodology, portability metrics) are designed to be tool-landscape-independent, but the specific case-study numbers are not.

\textit{External validity: site specificity.} The study is embedded in a regulated contracting context at a single organization operating under Zero Trust governance. Artifact quality standards (contract deliverable acceptance, regulated-environment curriculum compliance) reflect that context. Practitioners operating in commercial, academic, or nonprofit settings will face different acceptance criteria, different tool-access constraints (many commercial AI tools are restricted in regulated environments and vice versa), and different governance overhead floors. The directional findings should transfer, but the absolute numerical values should not be assumed portable to other organizational contexts without replication.

\textit{Construct validity: validator independence.} For the four no-prior-expertise domains validated chiefly through stakeholder acceptance (video, presentations, web, curriculum), quality validation is the least independent, since acceptance comes from parties with working relationships to the practitioner's organization. Advancement past desk-review to substantive peer review for academic publications and automated test pass rates for software represent the most independent validation signals available at the thresholds defined in \Cref{sec:metrics}. For academic publications, $Q_d$ is assessed against the desk-review-advancement threshold defined in \Cref{sec:metrics}; outcomes across the companion manuscripts were mixed, including major-revision decisions and desk rejections, so $Q_d = 1$ is not claimed uniformly across the academic-publishing portfolio, and no claim of eventual acceptance is made. Stakeholder briefing acceptance and contracting authority approval, while meaningful operational signals, come from parties who may benefit from the practitioner's success and cannot be treated as fully independent quality assessments. Future studies should incorporate blinded external review of cross-domain outputs to establish quality levels independent of stakeholder relationships.

\subsection{What Transfers and What Varies}\label{sec:disc-generalizability}

The central claim is that prompt and context engineering skills are portable across tool boundaries, not that any specific tool portfolio is. Different practitioners in different organizations will assemble different portfolios for different domains. The transferable elements are: (1) the six-phase orchestration methodology (general across all domains and tool sets); (2) the portability metrics framework (transfer velocity, cross-domain output quality, orchestration overhead, coverage breadth, applicable to any tool portfolio); (3) the three-discipline progression (general); and (4) the governance checkpoint principle at tool boundaries (general and domain-agnostic).

The context-dependent elements are: (1) the specific tools in the portfolio, which varies by organization, domain, and tool maturity; (2) the specific professional domains served, which varies by role and organizational need; (3) the transfer velocity and coverage breadth achieved, which depend on practitioner skill, tool landscape maturity, and integration ecosystem; and (4) the operational context (this case study is embedded in a regulated contracting organization, but the methodology is domain-agnostic and should apply equally to commercial, academic, or nonprofit settings).

The portability claim can be situated within established skill transfer theory. Thorndike and Woodworth's classical identical elements theory~\cite{thorndike1901influence} predicts that transfer occurs when the source and target tasks share common elements. Barnett and Ceci~\cite{barnett2002when} extended this into a taxonomy of far transfer, distinguishing dimensions such as knowledge domain, physical context, temporal context, and functional context along which source and target tasks may differ. Augment engineering skill transfer maps onto this taxonomy as follows: the knowledge domain changes (software development to video production), the physical context is constant (same workstation, same practitioner), the temporal context is near (skills applied within weeks of development), and the functional context shifts from execution to orchestration. Under Barnett and Ceci's framework, the transfer observed in this case study qualifies as moderate-to-far transfer along the knowledge domain axis but near transfer along temporal and physical axes. The shared procedural elements (structured input assembly, output validation against rubrics, iterative refinement) constitute the identical elements that Thorndike's theory predicts would enable transfer. This theoretical grounding suggests that augment engineering portability is not anomalous but rather a predictable consequence of shared procedural structure across tool interactions.

\subsection{Expertise as Amplification and Enablement}\label{sec:disc-expertise}

The relationship between domain expertise and augment engineering success requires careful disaggregation. Where the practitioner had deep domain expertise (cybersecurity, software engineering, contract proposal writing), augment engineering amplified existing capability. The practitioner already understood what secure software architecture required, what reviewers expected in submissions advancing past desk-review, what contracting authorities demanded in proposals. AI tools accelerated production; domain expertise ensured quality. The high output quality in these domains (100\% test pass rates, advancement past desk-review to substantive peer review, contracting authority approval) reflects both deep expertise and prompt/context engineering methodology. The two contributions cannot easily be separated, and overclaiming the role of the methodology alone would be misleading.

In contrast, where the practitioner had no specialist training (video production, presentation design, curriculum design, academic publishing, web deployment), augment engineering enabled domain entry. The practitioner produced professional-grade work products by applying prompt and context engineering skills to purpose-built tools that carry domain knowledge (HeyGen for video production, Gamma for presentation design, Vercel for web deployment). The orchestration skill came from the practitioner; the domain knowledge was encoded in the tool. This distinction defines the boundary of the methodology's applicability: augment engineering does not replace deep expertise for tasks requiring tacit knowledge, physical skill, or domain intuition that cannot be encoded as structured inputs. A prompt engineer operating HeyGen can produce professional training videos; they cannot direct a feature film. The methodology enables domain coverage for tool-mediated tasks where the tool provides the domain-specific capability and the practitioner provides orchestration methodology.

Domain entry into areas where the practitioner lacked prior training (video production, presentation design, web deployment, curriculum design, and academic publishing) is the mechanism the framework predicts, distinct from marginal productivity improvement in already-competent domains. To illustrate: a practitioner with zero video production training achieved professional video delivery in four hours of tool use. This timeline is more consistent with portable methodology enabling rapid domain entry than with general intelligence alone, though the single-practitioner design cannot rule out confounding factors such as the practitioner's overall technical aptitude or the relative simplicity of the tool-mediated tasks.

This amplification-versus-enablement asymmetry produces a testable prediction: augment engineering methodology should yield larger marginal gains for practitioners entering domains where they lack prior expertise than for practitioners working in domains where they are already proficient. This prediction aligns with a convergent finding across recent empirical studies. Brynjolfsson et al.~\cite{brynjolfsson2023generative} report that AI assistance produces a 15\% average productivity increase, with gains concentrated among less experienced workers. Noy and Zhang~\cite{noy2023experimental} find that ChatGPT use reduces task completion time by 40\% with an 18\% quality improvement, again with larger gains for lower-performing workers. Merali~\cite{merali2024scaling} derives economic scaling laws showing that each 10$\times$ increase in model compute reduces task time by 12.3\%, with gains four times larger for lower-skilled workers. Celis et al.~\cite{celis2025mathematical} formalize this phenomenon through a mathematical framework in which job success probability exhibits sharp phase transitions as subskill abilities cross critical thresholds, and combining workers with complementary subskills (the decision-level strength of humans with the action-level execution of AI) significantly outperforms either alone. This cross-study convergence on \textit{productivity compression}, where AI disproportionately benefits lower-baseline performers, is precisely what augment engineering predicts: the enablement function (domain entry via portable skills applied to AI tools) generates larger delta than the amplification function (marginal improvement in already-competent domains).

\subsection{Production Acceleration Evidence}\label{sec:disc-impact}

\Cref{fig:wrights-law} presents a Wright's Law learning curve fitted to 82 instrumentally tracked sub-deliverables produced during the study period (10 HeyGen/Gamma artifacts excluded for lack of per-artifact instrumentation). The power-law fit ($\hat{b} = 0.44$, $p < 0.01$) is consistent with production acceleration across domains, with phase-level medians of 30~hours per artifact during initial tool onboarding (Phase~1), 9~hours during cross-domain integration (Phase~2), and 10~hours during portfolio-scale orchestration (Phase~3). The observed acceleration is concentrated in the Phase~1 to Phase~2 transition (median 30~hours to 9~hours); Phase~2 and Phase~3 are statistically indistinguishable (Mann-Whitney $p = 0.69$), and the marginal Phase~3 uptick is consistent both with cumulative learning saturating once the context engineering methodology becomes routine and with the Phase~3 portfolio shifting toward higher-complexity cross-tool deliverables (academic papers, web deployment pipelines). The aggregate power law intentionally does not control for per-artifact variance in complexity. A statistically significant slope coupled with a low $R^2$ is the expected signature of a real central trend embedded in high per-observation variance: the fit captures the direction and rate of production acceleration, not the artifact-level noise from complexity heterogeneity. The Wright's Law result complements the Cochran-Armitage trend test reported in \Cref{sec:cs-longitudinal}, together providing the primary statistical evidence for the prompt-quality association and portfolio-wide production acceleration that the framework predicts.

\subsection{Observable Failure Modes}\label{sec:disc-failures}

Four failure modes have been observed or anticipated:

\textit{Skill transfer overestimation.} The practitioner assumes prompt and context engineering skills transfer to a new tool, but the tool's input model differs from expectations: it does not accept structured text inputs, or it requires specialized domain knowledge that cannot be encoded as context. Mitigation: Phase 3 of the orchestration methodology (skill transfer assessment) explicitly evaluates transferability before integration design begins. A boundary case illustrates the point: Jira automation required traditional Python scripting (the audit rotation pipeline, \Cref{sec:cs-failures}), not prompt/context engineering skill transfer. The week-long transfer velocity for Jira reflects software development time, not meta-skill portability. Tools requiring programmatic configuration rather than structured natural-language input are not candidates for PE/CE transfer and should be categorized as instrumental dependencies rather than orchestrated capabilities.

\textit{Quality gate failure at tool boundaries.} Outputs from Tool A are passed to Tool B without adequate quality validation, propagating errors downstream. Mitigation: Phase 4 (integration design) requires explicit quality gates at every tool boundary, as described in \Cref{sec:governance}.

\textit{Terminal node limitation.} Some tools produce outputs that cannot serve as structured inputs to other tools, creating dead ends in cross-domain workflows. HeyGen video outputs exemplify this: they require separate hosting and manual embedding; they cannot be consumed programmatically by downstream tools. Integration design must account for terminal nodes and place them at the end of tool chains, not in the middle.

\textit{Orchestration overhead dominance.} The practitioner spends more time coordinating between tools than producing within tools, negating the coverage breadth benefit. The orchestration overhead metric (\Cref{sec:metrics}) explicitly tracks this ratio, and portfolio optimization (Phase 6) targets overhead reduction as the primary scaling lever.

The irreducible $O_{\text{floor}} \approx 5\%$ observed in the simplest pipeline (\Cref{eq:oh-decomp}) has a structural analog in Amdahl's Law for parallel computing. Just as Amdahl's Law establishes that the serial fraction of a program limits achievable speedup regardless of the number of processors, the human quality-gate review at tool boundaries constitutes an irreducible serial fraction of orchestration overhead that cannot be automated away:
\begin{equation}\label{eq:amdahl}
  O_h \geq O_{\text{floor}} = \frac{t_{\text{human-gate}}}{t_{\text{human-gate}} + t_{\text{prod}}}
\end{equation}
where $t_{\text{human-gate}}$ is the minimum time required for human validation at tool boundaries. No amount of automation ($O_{\text{auto}}$) can reduce $O_h$ below this floor, because the governance-first architecture (\Cref{sec:disc-governance}) requires human judgment at every trust boundary. Chiodo et al.~\cite{chiodo2025hitl} formalize this constraint through computational reductions, showing that effective human-in-the-loop oversight requires adequate understanding of the system, self-control to act on judgment, power to intervene, and aligned intentions, conditions that cannot be delegated to automation without losing the verification guarantee. Shen and Tamkin~\cite{shen2026skill} provide complementary empirical evidence: in a randomized experiment, developers who fully delegated coding tasks to AI showed some productivity gains but measurably weaker code comprehension and debugging ability, suggesting that the human oversight floor is not merely a governance requirement but a prerequisite for maintaining the practitioner's ability to orchestrate effectively. This is not a limitation to be overcome but a design constraint to be respected: the floor exists because human oversight at tool boundaries is what distinguishes augment engineering from fully autonomous pipelines and ensures both output quality and practitioner skill preservation~\cite{shen2026skill, chiodo2025hitl}. Practitioners should optimize toward the floor rather than attempting to eliminate it.

\subsection{Governance-First Architecture and the Trust Hierarchy}\label{sec:disc-governance}

The governance checkpoints described in \Cref{sec:governance} are not merely operational safeguards; they instantiate at the orchestration level the trust-redirection principle from this author's broader research program on specification-layer governance~\cite{calboreanu2026mandate}, authorization-layer governance~\cite{calboreanu2026lattice}, and execution-layer governance~\cite{calboreanu2026trace}. Augment engineering adds the orchestration level: do we trust this multi-tool pipeline? The irreducible governance overhead ($O_{\text{floor}} \approx 5\%$ from \Cref{eq:oh-decomp}) is a design property to respect, not a limitation to overcome (\Cref{sec:disc-failures} discusses why). The four governance frameworks were developed by the present author across a connected research program; readers should consider this when evaluating the trust hierarchy as independent validation. The broader principle that assurance attaches to a verifiable governance process rather than to any single component is not unique to this program; it is consistent with the risk-management posture of the NIST AI Risk Management Framework~\cite{nist-ai-rmf}, which locates AI assurance in governance functions rather than in properties of an individual system.

\subsection{Codification for Independent Replication}\label{sec:disc-codification}

The augment engineering methodology has been codified into a four-day instructional sequence intended to test whether the six-phase orchestration methodology and the portability metrics framework transfer to practitioners from diverse professional backgrounds. This codification follows the same competency-based progression observed in the case study: interaction-level optimization, then structured input pipeline design, then multi-tool orchestration. Empirical evaluation of learner outcomes is a separate planned study and is not claimed here.

\subsection{Directions for Future Replication}\label{sec:disc-future}

Six research directions emerge from the limitations and scope of this work:

\textit{Multi-practitioner validation (highest priority).} Apply the methodology with practitioners trained in prompt and context engineering methodology from diverse professional backgrounds. Measuring whether practitioners with different expertise profiles achieve comparable skill transfer velocity and coverage breadth expansion is the most direct test of the framework's central claim and the most critical step toward independent replication.

\textit{Controlled comparison.} Design a controlled experiment comparing augmented practitioners (trained in prompt/context/augment engineering) against non-augmented practitioners on cross-domain task batteries. Measure transfer velocity, output quality, orchestration overhead, and coverage breadth. This would establish whether the methodology produces measurable advantages beyond what general training in AI tool use would provide.

\textit{Standardized portability benchmark.} Develop a benchmark suite that measures transfer velocity, cross-domain output quality, orchestration overhead, and coverage breadth in a controlled setting, enabling direct comparison across studies and practitioners.

\textit{Disentangling practitioner learning from tool improvement.} The transfer velocity model (\Cref{eq:tv}) attributes $T_v$ reduction to cumulative practitioner skill. Yet Merali~\cite{merali2024scaling} demonstrates that exogenous tool improvement also reduces task time: each 10$\times$ increase in model compute yields an independent 12.3\% productivity gain. Over the five-month study period, the underlying LLMs themselves improved (Claude model updates, ChatGPT capability expansions). Future studies should decompose $T_v$ into practitioner-learning and tool-improvement components, for instance by holding model version constant across adoption events or by measuring transfer velocity for tools adopted simultaneously on different model generations.

\textit{Tool landscape evolution tracking.} As AI tools emerge, mature, and retire, document how practitioner portfolios evolve. Which tools enter the portfolio, which exit, and what factors drive those decisions? This longitudinal tracking would clarify the relationship between tool ecosystem dynamics and practitioner capability.

\textit{Organizational scaling.} Study how augment engineering scales from individual practitioners to teams. When multiple augment engineers operate in the same organization, how are portfolios coordinated? What governance standards are necessary to maintain quality across practitioners? This addresses the transition from formative case study to organizational deployment.

\section{Conclusion}\label{sec:conclusion}

Organizations now deploy diverse AI tools across professional domains, facing a critical challenge with no settled answer: who operates these tools, and how? Classical staffing models assume that tool operation requires hiring domain-specialist staff. This paper proposes an alternative: augment engineering is the discipline of orchestrating multiple purpose-built AI tools across professional domains, enabled by the portability of prompt and context engineering skills across tool boundaries. When orchestration skill is portable and tools encode domain knowledge, a single practitioner can operate across multiple domains without being a specialist in each.

The augment engineering framework comprises three integrated components. The six-phase orchestration methodology (domain inventory, tool mapping, skill transfer assessment, integration design, orchestration execution, and portfolio optimization) provides a repeatable process from initial portfolio design through continuous improvement. The portability metrics framework (transfer velocity, cross-domain output quality, orchestration overhead, and coverage breadth) supplies quantitative measures for assessing capability expansion and identifying scaling bottlenecks. The governance checkpoint principle ensures that multi-tool orchestration does not sacrifice quality or accountability at tool boundaries through structured input specification, output validation, and documented decision rationale.

Formative empirical evidence comes from a five-month case study (November 2025 through March 2026) in which a single practitioner applied prompt and context engineering skills across an orchestration stack of five AI tools and five supporting infrastructure components spanning seven professional domains: software development, academic publishing, proposals, curriculum design, video production, presentation design, and web deployment. The practitioner produced professional-grade work products in domains where they had no specialist training (academic publishing, curriculum design, video production, presentation design, web deployment), demonstrating that the methodology enables genuine domain entry through skill transfer rather than merely improving productivity in already-competent domains.

Augment engineering completes a three-discipline progression. Prompt engineering enables interaction with a single tool for a single task. Context engineering structures inputs to enable reproducible behavior and knowledge transfer across repeated uses of the same tool. Augment engineering orchestrates a portfolio of tools across multiple domains, enabled by the portability of prompt and context engineering skills at tool boundaries. This progression is not merely cumulative; each level changes the operating assumptions. Prompt engineers work within a tool's design space; context engineers design that space through structured specification; augment engineers design the space between tools through governance checkpoints and orchestration methodology.

Returning to the research questions posed in \Cref{sec:introduction}: RQ1 asked whether prompt and context engineering skills are portable across tool and domain boundaries. The case-study observations are consistent with cross-tool skill transfer: transfer-velocity estimates decrease with successive tool adoptions (directional, not statistically powered), input-structuring methodology transferred architecturally intact across the five AI tools in the portfolio (the portability claim is limited to this subset; the five infrastructure components participate in the orchestration stack but follow traditional learning curves, \Cref{sec:cs-failures}), and the practitioner produced validated outputs under domain-appropriate thresholds (\Cref{sec:metrics}) in five domains without prior specialist training. Statistical confirmation requires multi-practitioner replication. RQ2 asked whether a systematic orchestration methodology can be defined and codified. The six-phase methodology was extracted from observed practice, formalized with defined inputs, outputs, and completion criteria for each phase, and illustrated through four distinct cross-domain workflows with measured portability metrics.

The central contribution is methodological: it describes how portable prompt and context engineering skills enable a reproducible orchestration pattern across multiple domains. The scope of that pattern is bounded by the practitioner's orchestration skill and by the boundaries documented in \Cref{sec:cs-failures}. The methodology has been codified into a teachable progression; empirical replication of transfer effectiveness across diverse practitioners is the immediate next research priority.


\section*{Acknowledgments}

This research was conducted at the Swift North AI Lab, a division of The Swift Group, LLC. The author acknowledges operational support from Swift North staff during the study period.

\section*{CRediT author statement}

Elias Calboreanu (sole author): Conceptualization, Methodology, Software, Validation, Formal Analysis, Investigation, Data Curation, Writing -- Original Draft, Writing -- Review \& Editing, Visualization, Project Administration. The author conceived the augment engineering framework, designed and operated the case study environment across all seven professional domains, compiled all quantitative and qualitative data, performed the statistical analyses, and wrote the manuscript.

\section*{Declaration of competing interest}

The author declares the following financial interests/personal relationships which may be considered as potential competing interests: the author is employed by The Swift Group, LLC, which operates the Swift North AI Lab described in the case study. The author is simultaneously the framework designer and the case study practitioner, which creates an inherent self-validation risk. The single-practitioner, single-site study design is a direct consequence of this dual role. To mitigate self-reporting bias, the paper relies on instrumentally collected metrics (automated test pass rates, audit convergence data, version control logs) and externally validated quality signals (advancement past desk-review to substantive peer review, contracting authority acceptance, production deployment) wherever possible. Independent multi-practitioner replication is explicitly identified as the highest-priority future work item (\Cref{sec:disc-future}). References~\cite{calboreanu2026context, calboreanu2026lattice, calboreanu2026mandate, calboreanu2026trace} are the author's own prior work and form the conceptual foundation for this paper. Readers should consider this affiliation when interpreting results.

\section*{Funding}

No funding was received to assist with the preparation of this manuscript. The work was conducted with internal operational support from The Swift Group, LLC; no external grant funding was received.

\section*{Ethical approval}

Not applicable. This study involved analysis of the author's own professional AI interactions and tool portfolio and did not involve human or animal subjects. As a self-study of professional practice, this research falls outside the scope of institutional review board (IRB) oversight. Future multi-practitioner replication studies will require IRB review and informed consent protocols.

\section*{Data availability}

Aggregated output metrics and tool-portfolio documentation are reported in the paper. The two coded datasets underlying the quantitative analyses, \fname{artifact\_manifest.csv}, listing the 82 instrumentally tracked artifacts used in the Wright's Law analysis (\Cref{fig:wrights-law}) with per-artifact domain category, primary tool, first-commit or first-deployment timestamp, chronological sequence index, estimated hours, phase assignment, and brief complexity notes (the 10 HeyGen/Gamma artifacts excluded from the quantitative fit are listed separately with an exclusion-rationale flag); and \fname{interaction\_coding.csv}, the 200-interaction coding sheet used for \Cref{fig:prompt-quality} and the Cochran-Armitage trend test, with columns for interaction ID, platform, date, prompt-sophistication level (L1--L4 per the rubric in \ref{app:rubric}), first-pass acceptance, iteration count to acceptance, and context-file count, are available from the corresponding author on reasonable request. These artifacts are not publicly released: they were produced in operational contexts for defense and commercial clients and are subject to client-confidentiality constraints. The coded extraction is nonetheless sufficient to derive every numerical claim in the paper. The 200-interaction corpus is the same corpus reported in the companion context engineering study~\cite{calboreanu2026context}; no additional interactions were coded for the present analysis. The Python scripts that generate the figures and reproduce the Cochran-Armitage, Kruskal-Wallis, and chi-square statistics are available from the corresponding author on the same basis. Raw operational data from the case-study environment is retained internally and is likewise available from the corresponding author on reasonable request, subject to proprietary and operational-security constraints.

\section*{Declaration of generative AI and AI-assisted technologies in the writing process}

During the preparation of this work the author used generative AI and AI-assisted technologies, including Anthropic Claude and OpenAI ChatGPT, in the following roles: assisting with drafting and revision of manuscript text; copy-editing; LaTeX template migration and bibliography hygiene; and supporting the multi-LLM audit cycles that informed manuscript revision. All substantive intellectual contributions, the augment engineering definition, the six-phase methodology, the portability metrics, the case-study design and data collection, and all analytical claims, are the sole work of the author. After using these tools, the author reviewed and edited all generated content and takes full responsibility for the content of the publication.

\appendix

\section{Prompt-Sophistication Rubric (L1--L4)}\label{app:rubric}

All 200 interactions were coded by the single practitioner using the following four-level scale. No inter-rater reliability assessment was performed; this is acknowledged as a limitation (\Cref{sec:disc-limitations}).

\begin{description}
\item[L1 -- Basic task articulation.] A single natural-language instruction with no explicit structure, no context files, and no output constraints. Example (anonymized): \textit{``Write a summary of the meeting notes.''}
\item[L2 -- Initial structure.] The prompt includes at least one structural element: a numbered instruction list, a role assignment, or a single context file. Example: \textit{``You are a technical writer. Summarize the attached meeting notes into three sections: decisions, action items, open questions.''}
\item[L3 -- Schema-enforced structure.] The prompt includes a structured context package (2+ files with distinct functional roles: authority document, source material, output constraints) and specifies output format or rubric. Example: \textit{``Using the attached requirements document (authority) and the prior draft (source), revise Section 3 to satisfy the compliance matrix (constraints). Output in LaTeX format matching the journal template.''}
\item[L4 -- Programmatic and infrastructure-integrated.] The prompt is part of an automated or semi-automated pipeline: schema-validated inputs, CI/CD-triggered execution, or cross-tool handoff with structured intermediaries. Example: \textit{``[Jira-triggered] Generate the audit report for lane 4 using the governance specification (authority), the latest test results (source), and the compliance rubric (constraints). Output JSON conforming to the audit schema.''}
\end{description}

\bibliographystyle{elsarticle-num}
\bibliography{references}

@misc{calboreanu2026mandate,
  author       = {Calboreanu, Elias},
  title        = {{MANDATE}: Multi-Agent Nominal Decomposition for Autonomous Task Execution},
  year         = {2026},
  month        = feb,
  howpublished = {SSRN},
  doi          = {10.2139/ssrn.6170328},
  note         = {\url{https://ssrn.com/abstract=6170328}}
}

@misc{calboreanu2026lattice,
  author       = {Calboreanu, Elias},
  title        = {{LATTICE}: Layered Architecture for Trusted and Transparent Intelligence
                  in Constrained Environments},
  year         = {2026},
  month        = jan,
  howpublished = {SSRN},
  doi          = {10.2139/ssrn.6151128},
  note         = {\url{https://ssrn.com/abstract=6151128}}
}

@misc{calboreanu2026trace,
  author       = {Calboreanu, Elias},
  title        = {{TRACE}: Trusted Runtime for Autonomous Containment and Evidence},
  year         = {2026},
  month        = feb,
  howpublished = {SSRN},
  doi          = {10.2139/ssrn.6212818},
  note         = {\url{https://ssrn.com/abstract=6212818}}
}

@techreport{calboreanu2026context,
  author      = {Calboreanu, Elias},
  title       = {Context Engineering: A Methodology for Structured Human-{AI} Collaboration},
  institution = {Capitol Technology University},
  type        = {Working Paper},
  number      = {v3.1},
  year        = {2026},
  month       = apr,
  note        = {Preprint: arXiv:2604.04258 (2026).
                 ORCID: \url{https://orcid.org/0009-0008-9194-0589}.
                 }
}

@techreport{calboreanu2026closedloop,
  author      = {Calboreanu, Elias},
  title       = {Closed-Loop Autonomous Software Development via Jira-Integrated
                 Backlog Orchestration},
  institution = {Swift North AI Lab},
  year        = {2026},
  note        = {In preparation, targeting the \textit{Automated Software Engineering} (Springer) special issue, 2026.
                 Preprint available from the corresponding author upon request}
}

@inproceedings{amershi2019guidelines,
  author    = {Amershi, Saleema and Weld, Dan and Vorvoreanu, Mihaela and others},
  title     = {Guidelines for Human-{AI} Interaction},
  booktitle = {Proceedings of the 2019 {CHI} Conference on Human Factors in Computing Systems},
  year      = {2019},
  doi       = {10.1145/3290605.3300233}
}

@article{bansal2021does,
  author  = {Bansal, Gagan and Wu, Tongshuang and Zhou, Joyce and others},
  title   = {Does the Whole Exceed its Parts? The Effect of {AI} Explanations on Complementary Team Performance},
  journal = {Proceedings of the 2021 CHI Conference on Human Factors in Computing Systems},
  year    = {2021},
  doi     = {10.1145/3411764.3445717}
}

@article{peng2023impact,
  author  = {Peng, Sida and Kalliamvakou, Eirini and Cihon, Peter and Demirer, Mert},
  title   = {The Impact of {AI} on Developer Productivity: Evidence from {GitHub Copilot}},
  journal = {arXiv preprint arXiv:2302.06590},
  year    = {2023}
}

@article{brynjolfsson2023generative,
  author  = {Brynjolfsson, Erik and Li, Danielle and Raymond, Lindsey R.},
  title   = {Generative {AI} at Work},
  journal = {Quarterly Journal of Economics},
  volume  = {140},
  number  = {2},
  pages   = {889--942},
  year    = {2025},
  doi     = {10.1093/qje/qjae044}
}

@article{autor2015there,
  author  = {Autor, David H.},
  title   = {Why Are There Still So Many Jobs? The History and Future of Workplace Automation},
  journal = {Journal of Economic Perspectives},
  volume  = {29},
  number  = {3},
  pages   = {3--30},
  year    = {2015}
}

@article{acemoglu2019automation,
  author  = {Acemoglu, Daron and Restrepo, Pascual},
  title   = {Automation and New Tasks: How Technology Displaces and Reinstates Labor},
  journal = {Journal of Economic Perspectives},
  volume  = {33},
  number  = {2},
  pages   = {3--30},
  year    = {2019}
}

@article{white2023prompt,
  author  = {White, Jules and Fu, Quchen and Hays, Sam and others},
  title   = {A Prompt Pattern Catalog to Enhance Prompt Engineering with {ChatGPT}},
  journal = {arXiv preprint arXiv:2302.11382},
  year    = {2023}
}

@article{reynolds2021prompt,
  author  = {Reynolds, Laria and McDonell, Kyle},
  title   = {Prompt Programming for Large Language Models: Beyond the Few-Shot Paradigm},
  journal = {arXiv preprint arXiv:2102.07350},
  year    = {2021}
}

@article{wei2022chain,
  author  = {Wei, Jason and Wang, Xuezhi and Schuurmans, Dale and others},
  title   = {Chain-of-Thought Prompting Elicits Reasoning in Large Language Models},
  journal = {Advances in Neural Information Processing Systems},
  volume  = {35},
  year    = {2022}
}

@misc{langchain2023,
  author       = {Chase, Harrison},
  title        = {{LangChain}: Building Applications with {LLMs} through Composability},
  year         = {2023},
  howpublished = {GitHub repository},
  note         = {\url{https://github.com/langchain-ai/langchain}}
}

@misc{autogpt2023,
  author       = {Richards, Toran Bruce},
  title        = {{Auto-GPT}: An Autonomous {GPT-4} Experiment},
  year         = {2023},
  howpublished = {GitHub repository},
  note         = {\url{https://github.com/Significant-Gravitas/Auto-GPT}}
}

@misc{crewai2024,
  author       = {Moura, Jo\~{a}o},
  title        = {{CrewAI}: Framework for Orchestrating Role-Playing, Autonomous {AI} Agents},
  year         = {2024},
  howpublished = {GitHub repository},
  note         = {\url{https://github.com/crewAIInc/crewAI}}
}

@article{hong2023metagpt,
  author  = {Hong, Sirui and Zhuge, Mingchen and Chen, Jonathan and others},
  title   = {{MetaGPT}: Meta Programming for a Multi-Agent Collaborative Framework},
  journal = {arXiv preprint arXiv:2308.00352},
  year    = {2023}
}

@article{yang2024sweagent,
  author  = {Yang, John and Jimenez, Carlos E. and Wettig, Alexander and others},
  title   = {{SWE}-agent: Agent-Computer Interfaces Enable Automated Software Engineering},
  journal = {arXiv preprint arXiv:2405.15793},
  year    = {2024}
}

@article{lai2023towards,
  author  = {Lai, Vivian and Chen, Chacha and Smith-Renner, Alison and others},
  title   = {Towards a Science of Human-{AI} Decision Making: An Overview of Design Space in Empirical Human-Subject Studies},
  journal = {Proceedings of the 2023 ACM Conference on Fairness, Accountability, and Transparency},
  year    = {2023},
  doi     = {10.1145/3593013.3594087}
}

@inproceedings{sahay2020supporting,
  author    = {Sahay, Apurvanand and Indamutsa, Arsene and Di Ruscio, Davide and Pierantonio, Alfonso},
  title     = {Supporting the Understanding and Comparison of Low-Code Development Platforms},
  booktitle = {Proceedings of the 46th Euromicro Conference on Software Engineering and Advanced Applications (SEAA)},
  year      = {2020},
  pages     = {171--178},
  doi       = {10.1109/SEAA51224.2020.00036}
}

@article{barnett2002when,
  author  = {Barnett, Susan M. and Ceci, Stephen J.},
  title   = {When and Where Do We Apply What We Learn? A Taxonomy for Far Transfer},
  journal = {Psychological Bulletin},
  volume  = {128},
  number  = {4},
  pages   = {612--637},
  year    = {2002},
  doi     = {10.1037/0033-2909.128.4.612}
}

@article{thorndike1901influence,
  author  = {Thorndike, Edward L. and Woodworth, Robert S.},
  title   = {The Influence of Improvement in One Mental Function upon the Efficiency of Other Functions},
  journal = {Psychological Review},
  volume  = {8},
  number  = {3},
  pages   = {247--261},
  year    = {1901}
}

@article{hutter2021learning,
  author  = {Hutter, Marcus},
  title   = {Learning Curve Theory},
  journal = {arXiv preprint arXiv:2102.04074},
  year    = {2021},
}

@article{viering2021shape,
  author  = {Viering, Tom and Loog, Marco},
  title   = {The Shape of Learning Curves: A Review},
  journal = {IEEE Transactions on Pattern Analysis and Machine Intelligence},
  volume  = {44},
  number  = {12},
  pages   = {9578--9597},
  year    = {2022},
  doi     = {10.1109/TPAMI.2021.3120763},
  note    = {arXiv:2103.10948}
}

@article{narayanan2025wrights,
  author  = {Narayanan, Rajesh P. and Pace, R. Kelley},
  title   = {Can the Nexus of Scaling Laws Coupled with Constant or Variable Elasticity
             of Substitution Predict {AI} and Other Technology Adoption?},
  journal = {arXiv preprint arXiv:2502.00909},
  year    = {2025},
}

@article{kim2025scaling,
  author  = {Kim, Yubin and Gu, Ken and Park, Chanwoo and others},
  title   = {Towards a Science of Scaling Agent Systems},
  journal = {arXiv preprint arXiv:2512.08296},
  year    = {2025},
}

@inproceedings{su2025difficulty,
  author    = {Su, Jinwei and others},
  title     = {Difficulty-Aware Agent Orchestration in {LLM}-Powered Workflows},
  booktitle = {arXiv preprint arXiv:2509.11079},
  year      = {2025},
}

@article{tan2022transferability,
  author  = {Tan, Yang and Li, Yang and Huang, Shao-Lun},
  title   = {Transferability-Guided Cross-Domain Cross-Task Transfer Learning},
  journal = {arXiv preprint arXiv:2207.05510},
  year    = {2022},
}

@article{fragiadakis2024evaluating,
  author  = {Fragiadakis, George and others},
  title   = {Evaluating Human-{AI} Collaboration: A Review and Methodological Framework},
  journal = {arXiv preprint arXiv:2407.19098},
  year    = {2024},
}

@article{shao2025future,
  author  = {Shao, Yijia and others},
  title   = {Future of Work with {AI} Agents: Auditing Automation and Augmentation Potential across the {U.S.} Workforce},
  journal = {arXiv preprint arXiv:2506.06576},
  year    = {2025},
}

@inproceedings{celis2025mathematical,
  author    = {Celis, L. Elisa and Huang, Lingxiao and Vishnoi, Nisheeth K.},
  title     = {A Mathematical Framework for {AI}-Human Integration in Work},
  booktitle = {Proceedings of the 42nd International Conference on Machine Learning (ICML)},
  pages     = {6978--7012},
  year      = {2025},
  series    = {PMLR},
  volume    = {267},
  note      = {arXiv:2505.23432}
}

@article{noy2023experimental,
  author  = {Noy, Shakked and Zhang, Whitney},
  title   = {Experimental Evidence on the Productivity Effects of Generative Artificial Intelligence},
  journal = {Science},
  volume  = {381},
  number  = {6654},
  pages   = {187--192},
  year    = {2023},
  doi     = {10.1126/science.adh2586}
}

@article{merali2024scaling,
  author  = {Merali, Ali},
  title   = {Scaling Laws for Economic Productivity: Experimental Evidence in {LLM}-Assisted Translation},
  journal = {arXiv preprint arXiv:2409.02391},
  year    = {2024},
}

@article{millinghoffer2025transfer,
  author  = {Millinghoffer, Andr\'{a}s and Bolg\'{a}r, Bence and Antal, P\'{e}ter},
  title   = {Characterization of Transfer Using Multi-task Learning Curves},
  journal = {arXiv preprint arXiv:2512.24866},
  year    = {2025},
}

@article{xu2026evolution,
  author  = {Xu, Haoyuan and others},
  title   = {The Evolution of Tool Use in {LLM} Agents: From Single-Tool Call to Multi-Tool Orchestration},
  journal = {arXiv preprint arXiv:2603.22862},
  year    = {2026},
}

@inproceedings{dang2025evolving,
  author    = {Dang, Yufan and Qian, Chen and others},
  title     = {Multi-Agent Collaboration via Evolving Orchestration},
  booktitle = {Advances in Neural Information Processing Systems (NeurIPS)},
  year      = {2025},
  note      = {arXiv:2505.19591}
}

@article{shen2026skill,
  author  = {Shen, Judy Hanwen and Tamkin, Alex},
  title   = {How {AI} Impacts Skill Formation},
  journal = {arXiv preprint arXiv:2601.20245},
  year    = {2026},
}

@article{chiodo2025hitl,
  author  = {Chiodo, Maurice and others},
  title   = {Formalising Human-in-the-Loop: Computational Reductions, Failure Modes, and Legal-Moral Responsibility},
  journal = {arXiv preprint arXiv:2505.10426},
  year    = {2025},
}

@article{vishnyakova2026context,
  author  = {Vishnyakova, Vera V.},
  title   = {Context Engineering: From Prompts to Corporate Multi-Agent Architecture},
  journal = {arXiv preprint arXiv:2603.09619},
  year    = {2026},
}

@techreport{nist-ai-rmf,
  author      = {{National Institute of Standards and Technology}},
  title       = {Artificial Intelligence Risk Management Framework ({AI RMF} 1.0)},
  institution = {NIST},
  type        = {Special Publication},
  number      = {100-1},
  year        = {2023},
  doi         = {10.6028/NIST.AI.100-1}
}

@techreport{chapman2010learning,
  author      = {Chapman, Bryan},
  title       = {How Long Does It Take to Create Learning?},
  institution = {Chapman Alliance},
  year        = {2010},
  note        = {Research study on e-learning and instructor-led training development ratios}
}

@article{wright1936factors,
  author  = {Wright, Theodore Paul},
  title   = {Factors Affecting the Cost of Airplanes},
  journal = {Journal of the Aeronautical Sciences},
  volume  = {3},
  number  = {4},
  pages   = {122--128},
  year    = {1936},
  doi     = {10.2514/8.155}
}

@book{wohlin2012experimentation,
  author    = {Wohlin, Claes and Runeson, Per and H{\"o}st, Martin and Ohlsson, Magnus C. and Regnell, Bj{\"o}rn and Wessl{\'e}n, Anders},
  title     = {Experimentation in Software Engineering},
  publisher = {Springer},
  address   = {Berlin, Heidelberg},
  year      = {2012},
  doi       = {10.1007/978-3-642-29044-2}
}

\end{document}